\documentclass[fleqn,usenatbib]{mnras}

\usepackage{newtxtext,newtxmath}
\usepackage{color}
\usepackage[T1]{fontenc}
\usepackage{siunitx}

\DeclareRobustCommand{\VAN}[3]{#2}
\let\VANthebibliography\thebibliography
\def\thebibliography{\DeclareRobustCommand{\VAN}[3]{##3}\VANthebibliography}

\usepackage{graphicx}	% Including figure files
\usepackage{amsmath}	% Advanced maths commands
\usepackage{pdflscape}

\newcommand{\XY}[2]{\left[\textrm{#1/#2}\right]}
\newcommand{\FeH}{\XY{Fe}{H}}

\newcommand{\kms}{km\,s$^{-1}$}
\newcommand{\Teff}{T_\textrm{eff}}
\newcommand{\logg}{\log g}

\newcommand{\vmic}{v_\textrm{mic}}

\title[High-resolution study of LMC EMP stars]{High-resolution spectroscopic study of extremely metal-poor stars in the Large Magellanic Cloud}

\author[W. S. Oh et al.]{
W. S. Oh,$^{1}$$^{,2}$\thanks{E-mail: weishen.oh@anu.edu.au}
T. Nordlander,$^{1}$$^{,2}$
G. S. Da Costa,$^{1}$$^{,2}$
M. S. Bessell$^{1}$$^{,2}$
and A. D. Mackey$^{1}$$^{,2}$
\\
$^{1}$Research School of Astronomy and Astrophysics, Australian National University, Canberra, ACT 2611, Australia\\
$^{2}$ARC Centre of Excellence for All Sky Astrophysics in 3 Dimensions (ASTRO 3D), Australia
}

\date{Accepted XXX. Received YYY; in original form ZZZ}

\pubyear{2023}

\begin{document}
\label{firstpage}
\pagerange{\pageref{firstpage}--\pageref{lastpage}}
\maketitle

\begin{abstract}
We present detailed abundance results based on UVES high dispersion spectra for 7 very and extremely metal-poor stars in the Large Magellanic Cloud.  We confirm that all 7 stars, two of which have [Fe/H] $\leq$ --3.0, are the most metal-poor stars discovered so far in the Magellanic Clouds. The element abundance ratios are generally consistent with Milky Way halo stars of similar [Fe/H] values. We find that 2 of the more metal-rich stars in our sample are enhanced in r-process elements. This result contrasts with the literature, where all nine metal-poor LMC stars with higher [Fe/H] values than our sample were found to be rich in r-process elements. The absence of r-process enrichment in stars with lower [Fe/H] values is consistent with a minimum delay timescale of $\sim$100 Myr for the neutron star binary merger process to generate substantial r-process enhancements in the LMC. We find that the occurrence rate of r-process enhancement (r-I or r-II) in our sample of very and extremely metal-poor stars is statistically indistinguishable from that found in the Milky Way's halo, although including stars from the literature sample hints at a larger r-II frequency the LMC. Overall, our results shed light on the earliest epochs of star formation in the LMC that may be applicable to other galaxies of LMC-like mass.\\

\end{abstract}

\begin{keywords}
stars:abundances -- stars: Population II -- Magellanic Clouds
\end{keywords}

%%%%%%%%%%%%%%%%%%%%%%%%%%%%%%%%%%%%%%%%%%%%%%%%%%

%%%%%%%%%%%%%%%%% BODY OF PAPER %%%%%%%%%%%%%%%%%%

\section{Introduction}

Extremely metal-poor (EMP) stars $(\rm[Fe/H] \leq -3.0)$ are some of the oldest stars that we can currently observe in the Universe. Although they are not the first-generation stars formed out of the original gas from the Big Bang, EMP stars still play a key role in terms of uncovering the properties of the first stars in the Universe. This is because they are second-generation stars that were formed out of gas enriched by the supernovae of their metal-free predecessors. EMP stars can help us understand the formation and evolution of galaxies during the early epochs of star formation, as they can reveal information about the properties of the first stars and the initial conditions of star formation in young galaxies \citep[e.g.][]{Frebel2015}.

Dwarf galaxies, which are smaller and less massive than their larger counterparts, serve as ideal laboratories for the study of EMP stars. Due to their relatively simple stellar populations, low metallicities and low stellar densities, dwarf galaxies provide an optimal environment for identifying and investigating these rare stars. However, the study of EMP stars in this environment has generally been quite limited due to observational challenges in terms of analysing stars in these distant objects (20-200 kpc) relative to the Milky Way. As such, EMP stars with metallicities down to only $\rm[Fe/H]=-4.11$ have been discovered in dwarf galaxies \citep{Skuladottir2021}. This is in contrast to the study of EMP stars in the Galactic halo, where the lowest ever detected iron abundance in a star (SMSS J160540.18--144323.1; $\rm[Fe/H] = -6.2$; \citealt{Nordlander2019}) and the most iron-poor star known (SMSS J031300.36--670839.3; $\rm[Fe/H] \leq -6.5$; \citealt{Keller2014}) have been found. Furthermore, EMP stars in dwarf galaxies have been found to show some significant differences in key elements (e.g. C, Na, and n-capture elements) compared to the Milky Way halo, which could be attributed to the lower masses \citep[e.g.][]{Tafelmeyer2010, Ishigaki2014, Mashonkina2017, Spite2018}. 

Ultra-faint dwarf galaxies (UFDs), which are smaller and less luminous compared to the Milky Way's other dwarf galaxy satellites, present another interesting environment to study EMP stars. While the metallicity range of the Milky Way halo and dwarf galaxies is comparable, there seems to be a notable absence of stars with $\rm[Fe/H] \geq -1.5$ in the UFDs \citep{Chiti2018}. Additionally, investigations of the Grus I and Triangulum II UFDs revealed a significant deficiency in neutron capture elements, leading to the inference that the primary source of these elements lies in infrequently occurring processes \citep{Ji2019}. In contrast, the UFD Reticulum II was found to be r-process enhanced (\citealt{Ji2016}; \citealt{Ji2023}), implying that the chemical enrichment and star-formation history timescales for these objects, compared to more luminous dwarfs and Milky Way-type galaxies, have not yet been adequately constrained.

The Large and Small Magellanic Clouds are two of the most prominent Milky Way satellites, but little is known about their EMP star populations despite their significance in various fields of galactic archaeology \citep[e.g.][]{Nidever2017}. Currently, the lowest metallicities found in the Magellanic Clouds are not as extreme as those observed in the Milky Way halo or dwarf satellites (e.g. \citealt{Reggiani2021}, \citealt{Oh2023}). Additionally, \citet{Reggiani2021} found that all of the metal-poor Magellanic stars in their sample, which range from $-2.4$ to $-1.5$ in $\FeH$, are r-process enhanced. In particular, their sample was found to display significant offsets ($\sim4\sigma$) in the r-process element europium relative to the Milky Way at comparable [Fe/H]. This has significant implications, suggesting that the Magellanic clouds experienced r-process enrichment events on timescales longer than core-collapse supernovae timescales but shorter than Type Ia supernovae timescales \citep{Reggiani2021}. This result might also be a consequence of the isolated evolution and prolonged history of accretion from the cosmic web of the Magellanic Clouds. This growing field of chemical abundance analysis has thus prompted a thorough analysis of EMP stars in the Magellanic Clouds.

In this paper, we conduct a high-resolution spectroscopic analysis with UVES on the ESO VLT of a sample of seven LMC stars that were initially selected based on SkyMapper photometry. Using a method similar to that of \citet{DaCosta2023}, the stars were followed up using low-resolution spectroscopy on the ANU 2.3m telescope, which produced a list of seven stars with metallicities $\rm [Fe/H] \leq -2.75$ \citep{Oh2023}. These stars constitute the most metal-poor stars so far discovered in this dwarf galaxy.

We present our sample selection, observations and data reduction in Section~\ref{sec:sec2}. In Section~\ref{sec:sec3}, we present the stellar parameter determination and the spectroscopic analysis methods for our EMP candidates. In Section~\ref{sec:sec4}, the abundance results of our measurements are presented and discussed, including interpreting the results in terms of nucleosynthesis timescales.

\section{Sample Selection, Observations and Data Reduction}
\label{sec:sec2}

\subsection{Sample selection}
\label{sec:selection} 

As described in \citet{Oh2023}, the targets were chosen based on SkyMapper photometry. Parallax and proper motion cuts (from Gaia), colour-magnitude cuts (by selecting the red giant branch region) and a metallicity-sensitive cut were also applied to the sample. The stars were subsequently observed at low resolution using the WiFeS integral field spectrograph at the ANU 2.3m telescope \citep{Oh2023}. Additional information regarding the photometric selection and WiFeS spectroscopy process can be found in \citet{DaCosta2019}. To determine estimates of effective temperature $(\Teff)$, surface gravity $(\logg)$, and metallicity $(\FeH)$, a spectrophotometric flux fitting method was applied to the WiFeS spectra, using the methodology first described in \citet{Bessell2007} and \citet{Norris2013}. The LMC membership of our candidates was also determined using their measured radial velocities \citep{Oh2023}. 

\subsection{Observations and Data Reduction}
\label{sec:observation} 

Our 7 best candidates (most metal-poor based on the WiFeS spectra) were then observed (Programme ID: 108.22D8) using the high-resolution Ultraviolet and Visual Echelle Spectrograph \citep[UVES;][]{Dekker2000} mounted on the Very Large Telescope (VLT) operated by the European Southern Observatory. The wavelength range, slit width and resolving power for the blue arm were 3289--4525 \AA, $1.0\arcsec$ and 40,000 respectively. For the red arm, the values were 4780--6801 \AA, $0.6\arcsec$ and 67,000. The spectroscopic observations were obtained over a few nights in service mode, spanning from August 2021 to January 2022. Each star was observed for 3 hours and 20 minutes in total, resulting in a final S/N per binned pixel of around 20 and 38 at 
3360~\AA\/ and 3900~\AA\/ respectively for the blue arm, and 46 and 70 at 5006~\AA\/ 
and 6296~\AA\/ respectively for the red arm.

We extracted the reduced data, processed with UVES pipeline version 5.10.13, from the ESO portal. The pipeline executes a comprehensive set of reduction procedures, encompassing bias correction, spectrum tracing, flat fielding, wavelength calibration as well as sky and cosmic ray removal. A single continuous spectrum was then generated for each star by coadding and merging multiple exposures. Finally, the spectra were continuum normalised by fitting high-order polynomials to the fluxes. Further information on the reduction and continuum normalisation process can be found in \citet{Yong2021}.

\section{Analysis}
\label{sec:sec3}

\subsection{Stellar parameters}
\label{sec:parameter} 
We first obtained the spectrophotometric temperatures of our LMC stars from \citet{Oh2023}. We then corrected these values by 50 K to account for the systematic offset between the spectrophotometric low-resolution analysis and accurate temperatures from Balmer lines and the infrared flux method (see \citet{Norris2013} and \citet{Yong2021} for more information).

Given that the EMP candidates are assumed to be in the LMC, we utilised the LMC distance to calculate fundamental $\logg$ values. This was calculated by using the canonical formula as shown in \citet{Oh2023}. The $\logg$ error was then estimated by assuming an error in mass of 0.2 $\rm{M_\odot}$, 100 K in $\rm{\Teff}$, 0.2 mag in distance modulus and 0.02 mag in E(B-V). Given that the maximum uncertainty contribution to $\logg$ from each term is $\sim0.05$ dex, we have decided to adopt a conservative uncertainty estimate of 0.3 dex.

\subsection{Spectroscopic analysis}
The set of model atmospheres utilised in this study was obtained from the $\alpha$-enhanced ($\rm[\alpha/Fe] = +0.4$), NEWODF grid of ATLAS9 models developed by \citet{Castelli2003}. We adopt solar abundances from \citet{Asplund2009}. The equivalent widths of a specific set of lines were determined for all the stars in our sample by fitting a Gaussian line profile according to the method described in \citet{McKenzie2022}. More information about the model atmosphere and the list of lines analysed in this work can be found in \citet{Yong2021}.

We used the MOOG LTE stellar line analysis program to determine the iron abundances using \ion{Fe}{I} and \ion{Fe}{II} lines. The microturbulent velocity, $\vmic$, was determined by ensuring that the abundances from \ion{Fe}{I} lines exhibit no trend relative to the reduced equivalent width, $\log(W _{\lambda}/\lambda)$.

Our method also compared the inferred metallicity from MOOG with the assumed metallicity used to generate the model atmosphere. If the difference between the two exceeded 0.2 dex, an updated model atmosphere using the new [Fe/H] value was computed. This was iterated until the derived stellar metallicities converged. Additionally, \ion{Fe}{I}lines that deviated from the median abundance by more than 0.5 dex or $3\sigma$ (based on the line-to-line scatter) were removed.

With the metallicities of our 7 stars obtained, we then computed the abundances for 24 elements. For Na, Mg, Al, Si, Ca, Sc,	Ti (\ion{Ti}{I} \& \ion{Ti}{II}), Cr, Mn, Co, Ni, Sr,	and Ba, this was done by measuring the equivalent widths of the lines and using MOOG to infer the abundances of each element. For the remaining 11 elements that were analysed, namely C (CH), N (CN), O, Zn, Zr, Y, La, Nd, Sm, Eu, and Dy, we measured the abundances by generating synthetic spectra via MOOG and adjusting their abundances until the synthetic spectra best fit the observed spectra. For Zn, Zr, Y, La, Nd, Sm, Eu, and Dy (which were not measured in \citealt{Yong2021}), the list of lines analysed was based on \citet{Sneden2003}.

Despite the potential effects of non-local thermodynamic equilibrium (NLTE) on abundance ratios, we opted to adopt LTE results for our analysis. This was due to the fact that our comparison studies also analysed their samples using LTE (\citealt{Jonsson2020}; \citealt{Yong2021}; \citealt{Reggiani2021}). While NLTE corrections can be crucial for accurately determining abundances in certain cases, we aimed to maintain consistency with the existing literature to allow a more meaningful comparison.

\subsection{Error analysis}

The statistical error was calculated by first estimating the uncertainty in the equivalent width measurement, using an analytic formula from \citet{Cayrel1988}:
\begin{equation}
    \sigma_{\rm{w}}=\frac{2.3\sqrt{\rm{w*p}}}{\rm{S/N}}
    \label{eq:statistical_error}
\end{equation}
where w is the line full width at half maximum (FWHM), p is the pixel wavelength step and S/N is the signal-to-noise ratio per pixel.
Then, we calculated the change in abundance measurements by re-running the abundance analysis using the equivalent width that was adjusted by the uncertainty value.

We also estimate the systematic impact on abundances from random errors in stellar parameters $(\Teff \pm 100\, \textrm{K}, \logg \pm \mathrm{0.3\, dex}, \vmic \pm 0.3\, \textrm{\kms}, \rm{[M/H] \pm 0.3})$, which are based on \citet{Oh2023}, \citet{DaCosta2023} and \citet{Yong2021}. The random uncertainty in the abundance and abundance ratios were then calculated by first perturbing, in turn, each adopted stellar parameter by its estimated uncertainty, and then combining the effects into a total random error via a sum of squares process. The typical values for the random errors are $\sim0.15$ dex.

We found that the statistical errors were, in general, at least an order of magnitude smaller than the random errors due to the uncertainties in stellar parameters and the line-to-line scatter. Hence, we only use and present the latter errors, including in the supplementary material provided with the online version.

\subsection{Radial velocities}

The radial velocities were determined by comparing the observed wavelengths of the lines (which come from the line centre of the fitted Gaussian profiles), for which equivalent widths were measured, with their corresponding rest wavelengths. The average standard deviation of these measurements was found to be 0.7 \kms. We also evaluated the zero point of our velocities by comparing our measurements with that of \citet{Oh2023} which measured radial velocities for the same stars but using the Ca II triplet region with medium-resolution ($R \approx 7,000$) spectra. The mean offset (measured minus medium-resolution reference) and standard deviation are $-1.2$ \kms\ and $10.8$ \kms\ respectively, indicating that the medium-resolution velocities used for membership selection are reliable with a small zero point uncertainty. Table~\ref{tab:stellar_param} provides the IDs, adopted atmospheric parameters and the radial velocities of our targets.

\section{Results and Discussion}
\label{sec:sec4}

\subsection{Stellar metallicities}
We have analysed high-resolution UVES spectra of 7 stars that are likely members of the LMC. We find metallicities in the range ${\rm [Fe/H]} = -2.49$ to $-3.13$, including two stars with $\rm[Fe/H] < -3$. The typical metallicity uncertainties are $\sim0.09$ dex. In comparison to our low-resolution metallicities, the offsets between the two types of measurements (high-resolution minus low-resolution) show a mean and standard deviation of 0.04 and 0.24 dex respectively. This is in line with the results from \citet{Oh2023} and \citet{DaCosta2019}, which state that the low-resolution metallicity measurements are precise to the 0.3 dex level. Thus, we can confirm that our medium-resolution selection reliably selects the most metal-poor stars, and all 7 stars are the most metal-poor stars analysed at high resolution in the LMC, and 2 are revealed as the first LMC EMP stars.

\begin{table*}
	\centering
	\caption{Coordinates and stellar parameters of the LMC very and extremely metal-poor stars.}
	\label{tab:stellar_param}
	\begin{tabular}{lccccccccccc} % xxx columns, alignment for each
		\hline
		SMSS DR3 & Gaia DR3 & RA & Dec.  & $g_{0}$ & $(g-i)_{0}$ & E(B-V) & $\Teff$ (K) & $\logg$ & $\vmic$ & [Fe/H] & rv (\kms)\\
		\hline
            500287810 & 4757943093811987072 & 05 28 29.6 & -61 26 49.6 & 16.223 & 1.096 & 0.050 & 4550& 0.95	& 2.40 & -2.49	& $283.7$\\
		500382880 & 4762709889033585024 & 05 24 24.1 & -59 16 05.5 & 16.456 & 1.001 & 0.032 & 4600 & 1.08 & 2.45 & -2.65 &	$268.3$\\
            497682788 & 4670107885871109760 & 04 03 47.0 & -64 30 55.7 & 16.557 & 0.969 & 0.047 & 4550 & 1.10 & 2.60 & -2.80 &	$233.3$\\
            471915910 & 5481017880523468800 & 06 08 14.9 & -62 07 23.5 & 16.600 & 0.948 & 0.049 & 4625 & 1.15	& 2.80 & -2.90	& $265.7$\\
            500766372 & 4762322654782102272 & 04 59 15.4 & -58 54 16.0 & 16.573 & 0.937 & 0.023 & 4650 & 1.17	& 2.75 & -2.90	& $234.6$\\
            497519424 & 4664660389875115392 & 04 54 53.6 & -64 07 52.2 & 16.616 & 1.001 & 0.036 & 4500 & 1.11	& 2.90 & -3.13	& $289.9$\\
            499901368 & 4727018642084375936 & 03 02 07.9 & -57 53 20.2 & 16.501 & 0.779 & 0.015 & 4750 & 1.23	& 1.80 & -3.13	& $326.2$\\
            \hline
	\end{tabular}
\end{table*}

\subsection{General comparison with the Milky Way halo}
Table \ref{tab:elements_table} and Figures \ref{fig:X_Fe} to \ref{fig:X_Fe_5} show the results of our abundance measurements compared to the results from a Milky Way halo sample (\citealt{Cayrel2004}; \citealt{Jacobson2015}; \citealt{Marino2019}; \citealt{Yong2021}) and a slightly more metal-rich sample in the LMC (\citealt{Jonsson2020}; \citealt{Reggiani2021}). For the \citet{Jonsson2020} sample, we applied the same cuts to the data as the ones for \citet{Oh2023} in terms of radius from LMC centre, parallax, proper motion and radial velocity. In this subsection, we will go through the abundance results in more detail. The abundance measurements of r-process elements will, however, be discussed in Section~\ref{sec:4.3} as they constitute the key results of our paper.

\subsubsection{Light Elements}
\label{sec:4.2.1}
The light elements measured in our study include carbon, nitrogen, oxygen, sodium and aluminium.

\begin{figure}
        \includegraphics[width=\columnwidth]{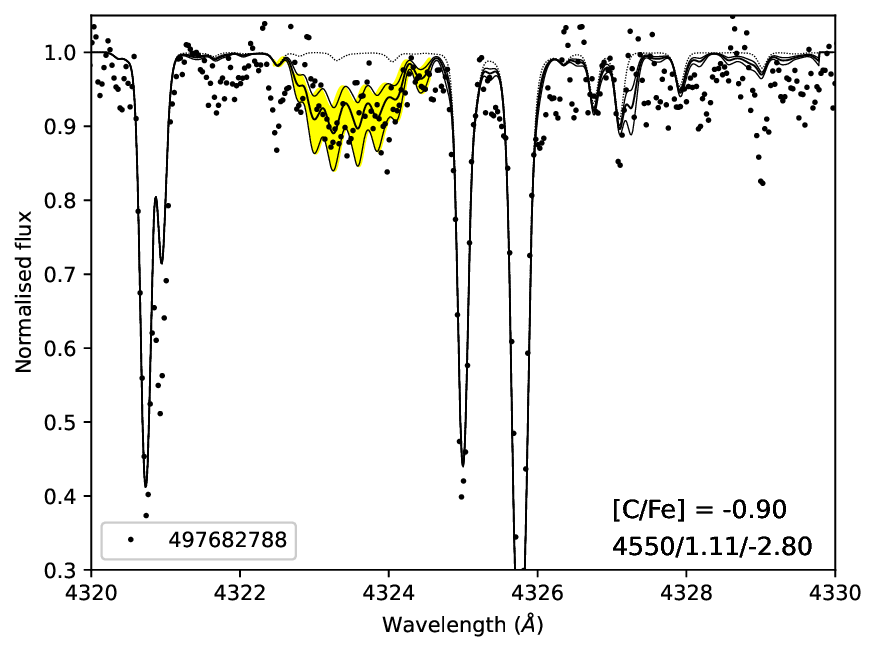}
        \includegraphics[width=\columnwidth]{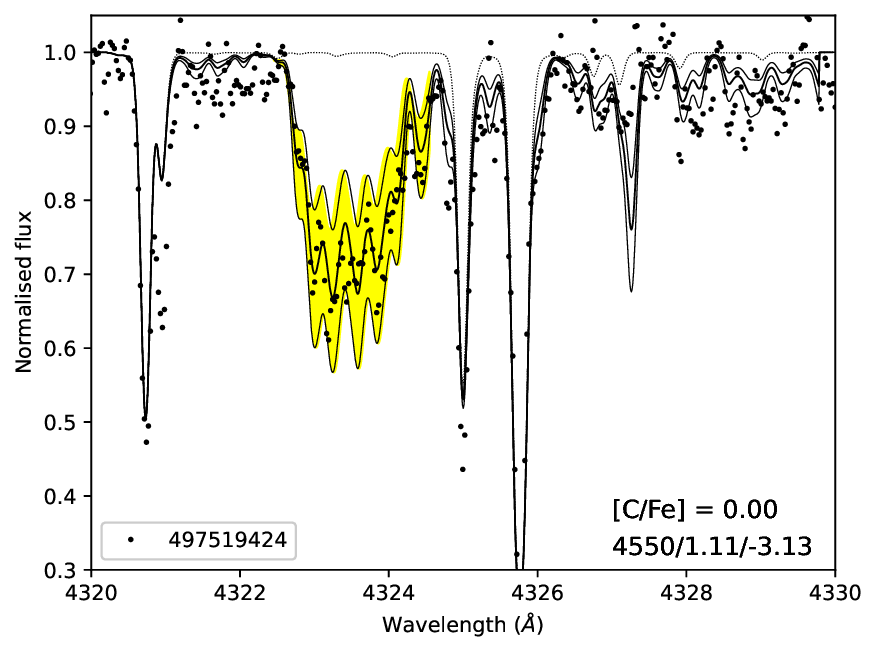}
    \caption{Comparison of observations (black dots) and synthetic spectra in the vicinity of the CH G-band at 4323 \AA\ for our least carbon abundant star (top: 497682788) and most carbon abundant star (bottom: 497519424). The synthetic spectra depicted by thin dotted lines correspond to $\rm{[C/Fe]} = -9$. The best-fitting synthetic spectra are illustrated by a thick black line, while the yellow shaded regions indicate a range of $\pm0.2$ dex around the best fit. The abundances have not been adjusted for evolutionary effects. The stellar parameters $\Teff/\logg/\FeH$ are shown.}
    \label{fig:C_Fe}
\end{figure}

We have successfully measured carbon abundances for all of our stars, and examples are shown in Fig.~\ref{fig:C_Fe}. Since our stars are located on the upper part of the RGB (beyond the bump), they have therefore likely undergone first dredge-up. With evolutionary corrections from \citet{Placco2014}, we can correct for this effect to recover estimates of their initial chemical composition. The typical correction value is $\sim+0.6$ dex. Our corrected carbon abundances have a mean and observed standard deviation of $\langle\rm{[C/Fe]\rangle} = 0.16 \pm 0.33$ dex, and span a range of $\rm [C/Fe] = -0.26$ to 0.63. Thus none of the sample meet the conventional definition of carbon-enhanced metal-poor stars ($\rm [C/Fe] > 0.7$, \citealt{Placco2014}). Interestingly, the 2 stars with the highest (uncorrected and corrected) [C/Fe] values are also the ones with the lowest [Fe/H] in our sample. As regards to the other 5 stars in our sample, these have slightly lower [C/Fe] values compared to MW stars with similar metallicities; the difference is $\sim0.5$ dex for the corrected abundances. Nevertheless, the values for these stars are similar to those of \citet{Jonsson2020} who find predominantly solar (corrected) [C/Fe] values in their LMC stars, which have metallicities down to $\rm [Fe/H] = -2$.  Our data extends this trend  to $\rm [Fe/H] \approx -3$, as shown in the upper left and upper right panels of Fig.~\ref{fig:X_Fe}.

\begin{figure}
        \includegraphics[width=\columnwidth]{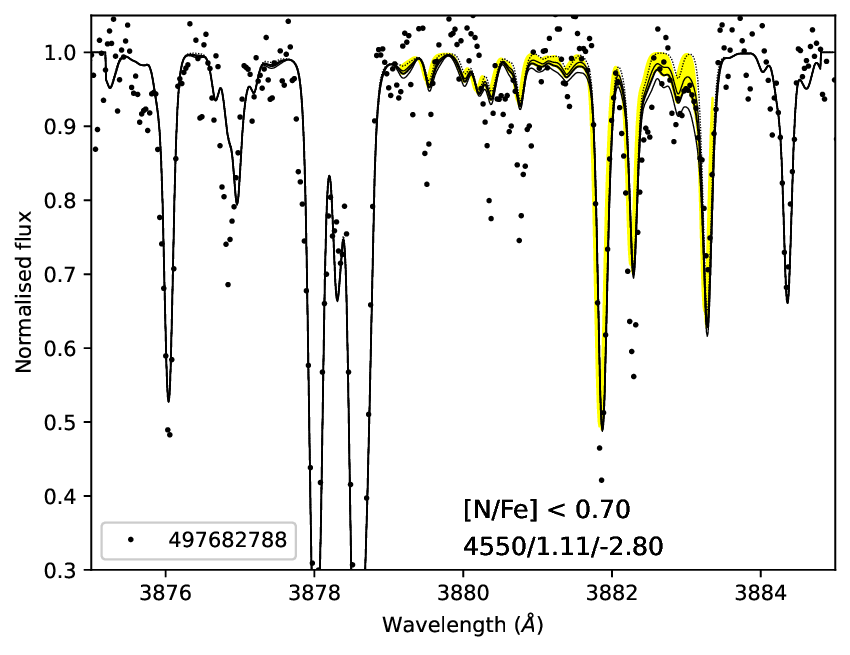}
        \includegraphics[width=\columnwidth]{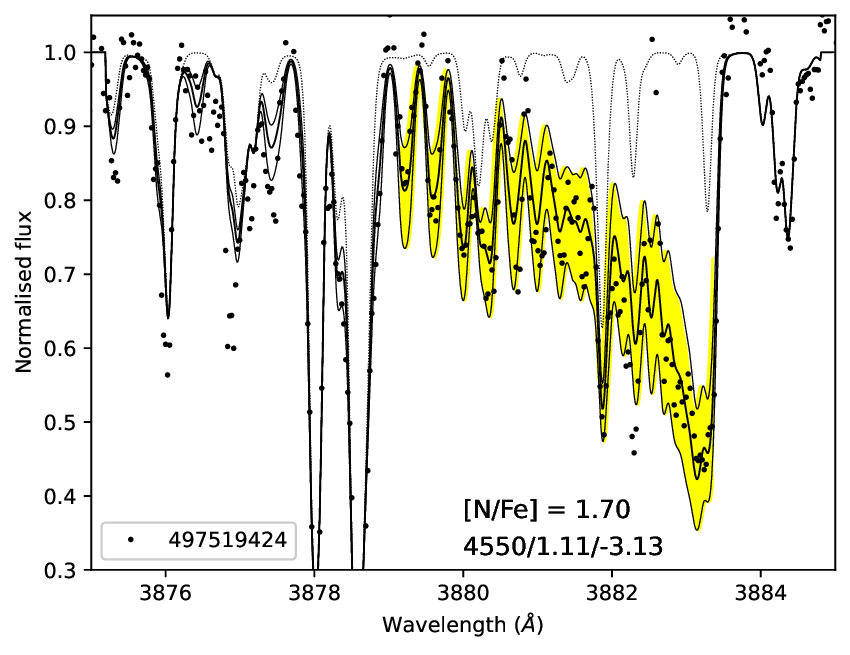}
    \caption{Same as Figure~\ref{fig:C_Fe} but for [N/Fe] illustrating the CN $3883$ \AA~feature.}
    \label{fig:N_Fe}
\end{figure}

For nitrogen, due to the high levels of noise in the near-UV part of the spectra, we were only able to definitively measure nitrogen for one star, 497519424, as shown in Fig.~\ref{fig:N_Fe}, using the CN bands at 3883 \AA. The upper limits for the other stars range from [N/Fe] $<$ 0.5 to $<$ 1.1.  Star 497519424 is strongly enhanced in nitrogen, $\rm{[N/Fe]=1.7} \pm 0.11$, relative to MW stars at similar [Fe/H] values and to those for LMC at higher metallicities. From \citet{Placco2014}, we can deduce that in such stars with much larger nitrogen abundances as compared to carbon, dredge-up causes [C/Fe] to become depleted as usual but [N/Fe] does not change much. Thus, we can confirm that this star has a high N abundance and a relatively low uncorrected C abundance ([C/Fe] = 0.0).  This makes it a Nitrogen Enhanced Metal Poor (NEMP) star as defined in the literature ($\rm{[N/Fe]} > 0.5$ and $\rm{[C/N]} < -0.5$; \citealt{Johnson2007}). This is thus the first NEMP star to be identified in the LMC. \citet{Yong2021} found a NEMP fraction of $55\% \pm 21\%$ for the stars in their sample in which [N/Fe] could be measured. Our fraction (one NEMP star from one [N/Fe] measurement) is consistent with that in \citet{Yong2021} but clearly a larger sample of [N/Fe] measurements in LMC very and extremely metal-poor stars is required to better establish the consistency, especially as our non-detections have upper limits that are near or even above the discriminant limit for nitrogen enhancement at [N/Fe] = 0.5. Nevertheless, the existence of a least one NEMP star in the LMC, along with those in the MW, shows that whatever process is responsible for the generation of such stars, it is evidently not dependent on the environment.

\begin{figure}
        \includegraphics[width=\columnwidth]{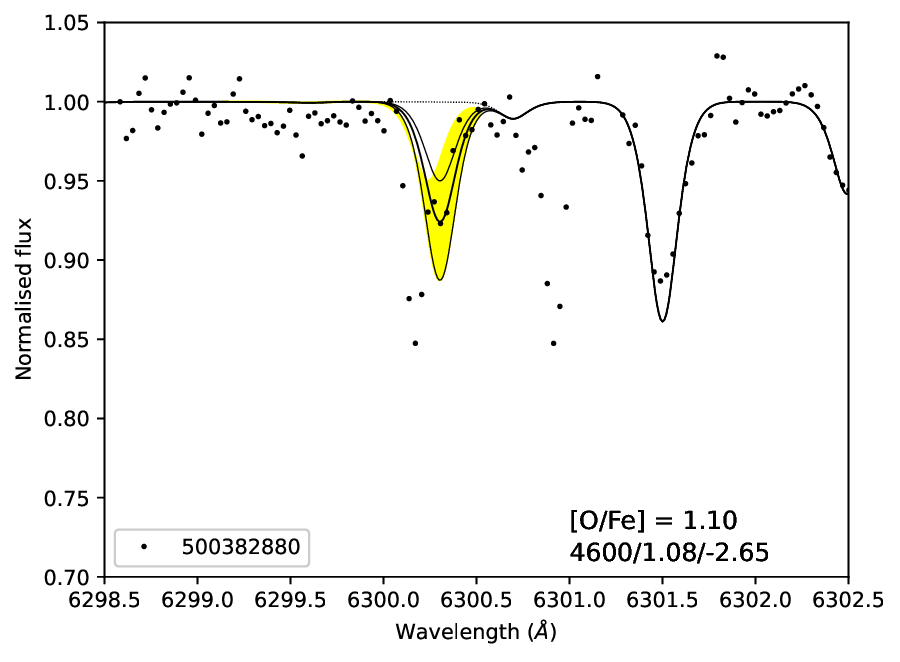}
    \caption{Same as Figure~\ref{fig:C_Fe} but for [O/Fe] illustrating the forbidden red line of neutral oxygen ([O I] $6300$ \AA).}
    \label{fig:O_Fe}
\end{figure}

For oxygen, the [\ion O I] 6300\,\AA\ feature was partially or fully affected by telluric features in the spectra for 4 of the 7 stars in our sample. Hence, we were only able to measure oxygen abundances for the 3 remaining stars. The oxygen spectra for the most O abundant star (500382880) is shown in Fig.~\ref{fig:O_Fe}. We found that our average oxygen abundance value ($\langle\rm{[O/Fe]\rangle=1.0 \pm 0.18}$) for the 3 stars with [O/Fe] determinations is slightly higher than that for the MW ($\rm{\langle[O/Fe]\rangle=0.69}$; \citealt{Cayrel2004}) across roughly the same [Fe/H] range. On the other hand, our average [O/Fe] value for these 3 stars is apparently much larger than the LMC value at higher metallicities ($\rm{\langle[O/Fe]\rangle=0.13}$; \citealt{Jonsson2020}).  We note, however,  that significant systematic differences may exist between measurements from the [\ion O I] line in this work and from the molecular near-infrared lines of OH in that work.

For sodium, the average abundance measurement for most of our stars ($\rm{\langle[Na/Fe]\rangle=0.10 \pm 0.11}$) is generally similar to those for the MW halo and the LMC at higher metallicities, and, with the exception of one star (the NEMP star 497519424 that has an exceptionally high [Na/Fe] value), there is no evidence for an intrinsic [Na/Fe] scatter for these stars. We note that the enhancement in [Na/Fe] in the NEMP star 497519424 exceeds that in any of the stars in \citet{Yong2021} or \citet{Reggiani2021}. This star will be further discussed in Section \ref{sec:4.4}.

For aluminium, our average abundance results ($\rm{\langle[Al/Fe]\rangle=-0.27 \pm 0.20}$) are generally higher than that of the MW halo, but similar to the LMC at higher metallicities. Given the size of the error bars for [Al/Fe], there does not seem to be any intrinsic scatter in our aluminium measurements.

\subsubsection{$\alpha$ Elements}
The elements magnesium, silicon, calcium and titanium are commonly denoted as $\alpha$ elements. This refers to elements that are primarily produced through nuclear fusion reactions involving successive $\alpha$-particle captures, such as stellar nucleosynthesis in core collapse supernovae. Our analysis shows that the measured abundances of the $\alpha$ elements are generally consistent with the observed $\alpha$ abundances in metal-poor stars within the Milky Way and in the LMC.

For magnesium, there is good agreement with the MW halo for 6 out of 7 stars in terms of our average abundance measurements ($\rm{\langle[Mg/Fe]\rangle=0.33 \pm 0.08}$). These stars extend to lower [Fe/H], the [Mg/Fe] vs [Fe/H] relation for the LMC seen in the results from \citet{Reggiani2021} and \citet{Jonsson2020}. One star, 500766372, appears Mg depleted with a low value of $\rm [Mg/Fe] = -0.02 \pm 0.09$, which is 0.3 dex below the average of the other stars in our sample.

For silicon, our average abundance and scatter ($\rm{\langle[Si/Fe]\rangle=0.47 \pm 0.31}$) are similar to that of the MW halo. The star-to-star scatter is significantly larger than the typical error bar, implying that the abundance dispersion is real. In addition, the average measurement of our 4 most silicon abundant stars ($\rm{\langle[Si/Fe]\rangle=0.68 \pm 0.25}$) are consistent with the high [Si/Fe] seen in the \citet{Reggiani2021} sample, while the average measurement for the other 3 ($\rm{\langle[Si/Fe]\rangle=0.20 \pm 0.08}$) are more consistent with the lower [Si/Fe] values seen in the \citet{Jonsson2020} sample at higher [Fe/H]. Interestingly, the silicon abundances from \citet{Reggiani2021}, like our results, have a large scatter, while the results from \citet{Jonsson2020} have a small scatter, even in the region where they overlap in metallicity. This could be due to the fact that, like us, \citet{Reggiani2021} used optical spectra, as opposed to the near-infrared H-band data of \citet{Jonsson2020}.

For calcium, our average abundance ($\rm{\langle[Ca/Fe]\rangle=0.25 \pm 0.09}$) shows excellent agreement with the MW halo and with most of the LMC stars from \citet{Reggiani2021} and \citet{Jonsson2020}. However, unlike the case for the other LMC samples, we do not see any evidence for a large scatter in [Ca/Fe] similar to that exhibited at higher [Fe/H] values.  Specifically, in our sample there are no stars with very high [Ca/Fe] values comparable to those seen in the \citet{Reggiani2021} results.  Additionally, at our [Fe/H] values, we see no stars with sub-solar [Ca/Fe] comparable to the results of \citet{Jonsson2020} at metallicities exceeding $\rm [Fe/H] = -2.0$.

For titanium, our abundance results ($\rm{\langle[\ion {Ti} I/Fe]\rangle=0.20 \pm 0.11}$; $\rm{\langle[\ion {Ti} {II}/Fe]\rangle=0.33 \pm 0.14}$) are similar to those for calcium, with good agreement with the MW halo and reasonable agreement with the LMC samples. There is no evidence for any intrinsic scatter in our [\ion {Ti} I/Fe] and [\ion {Ti} {II}/Fe] values, and there are no stars in our sample with [\ion {Ti} I/Fe] values comparable to the very high [\ion {Ti} I/Fe] measurement seen in one star in the \citet{Reggiani2021} sample.

\begin{figure*}
	\includegraphics[width=\columnwidth]{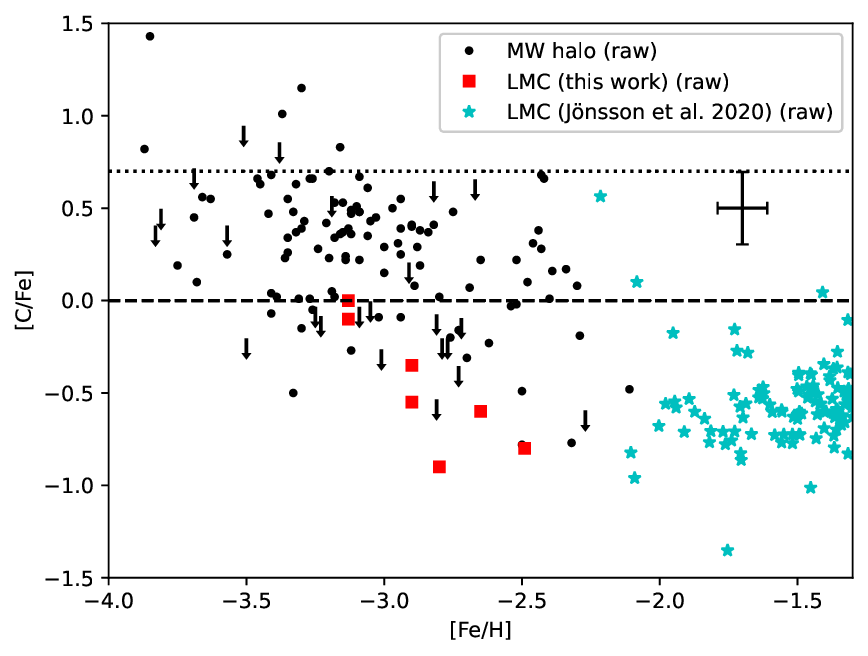}
        \includegraphics[width=\columnwidth]{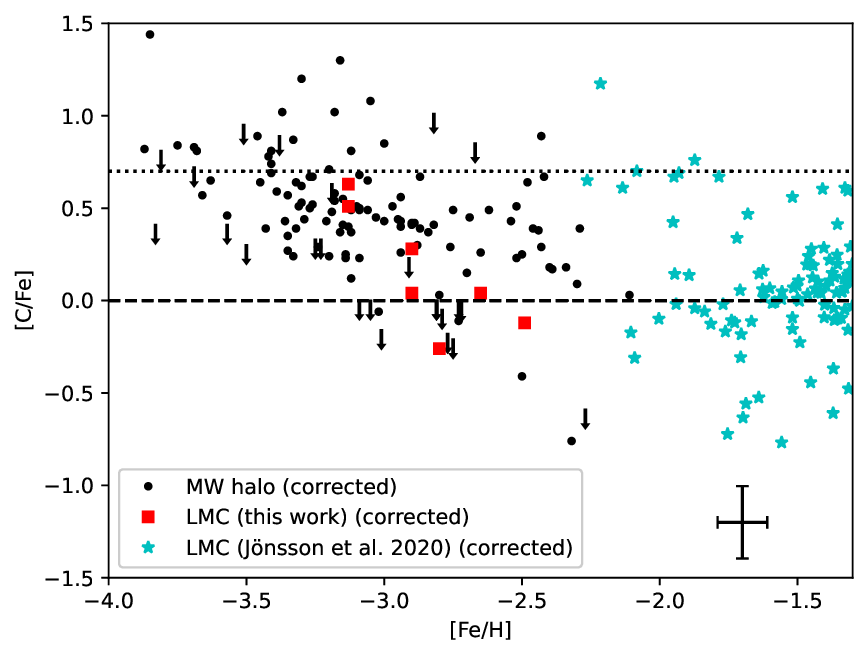}
        \includegraphics[width=\columnwidth]{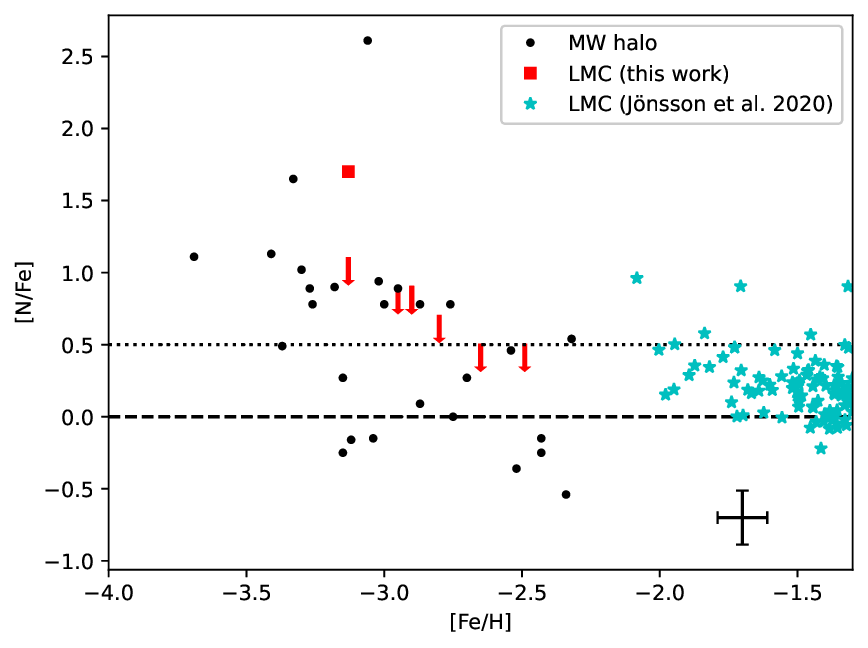}
        \includegraphics[width=\columnwidth]{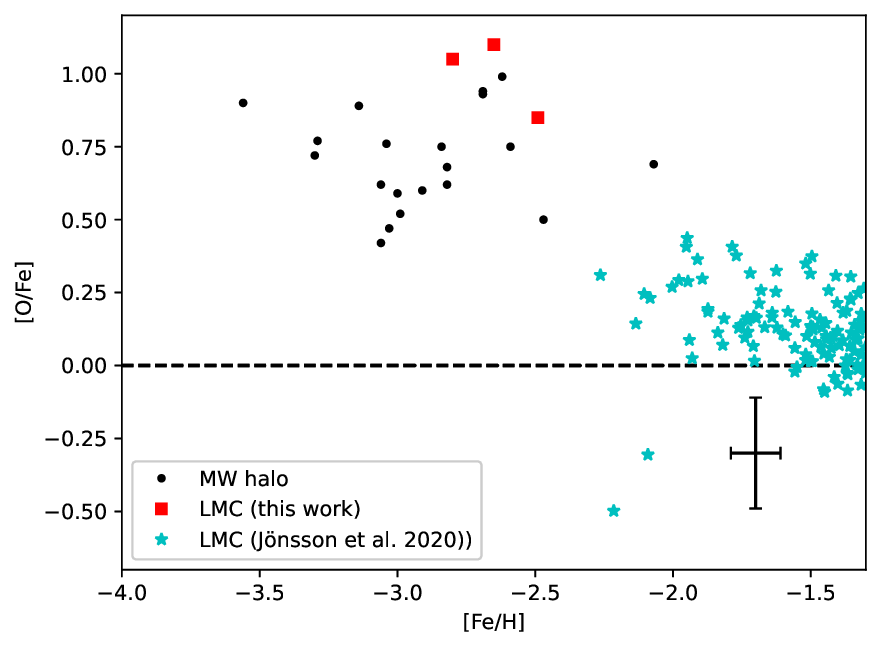}
        \includegraphics[width=\columnwidth]{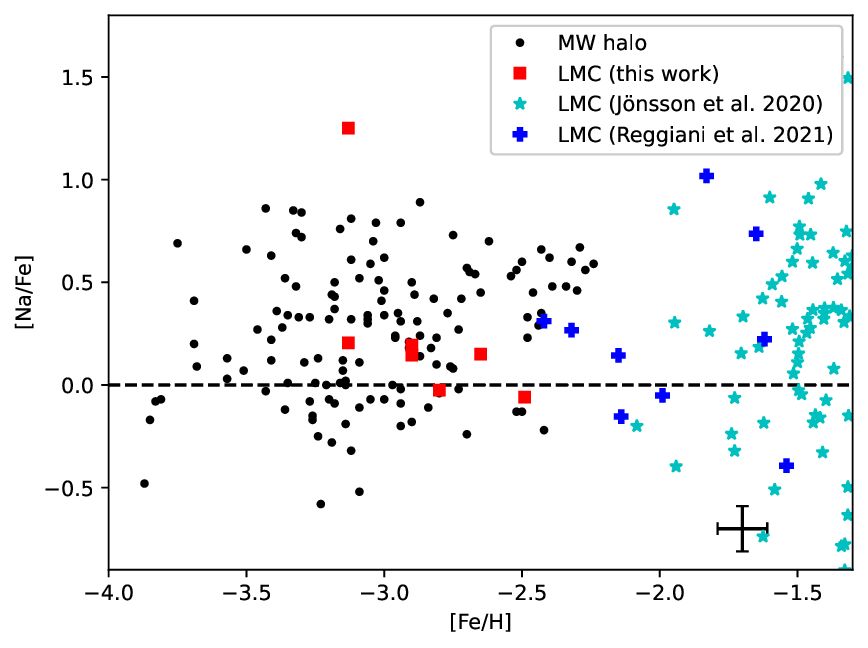}
        \includegraphics[width=\columnwidth]{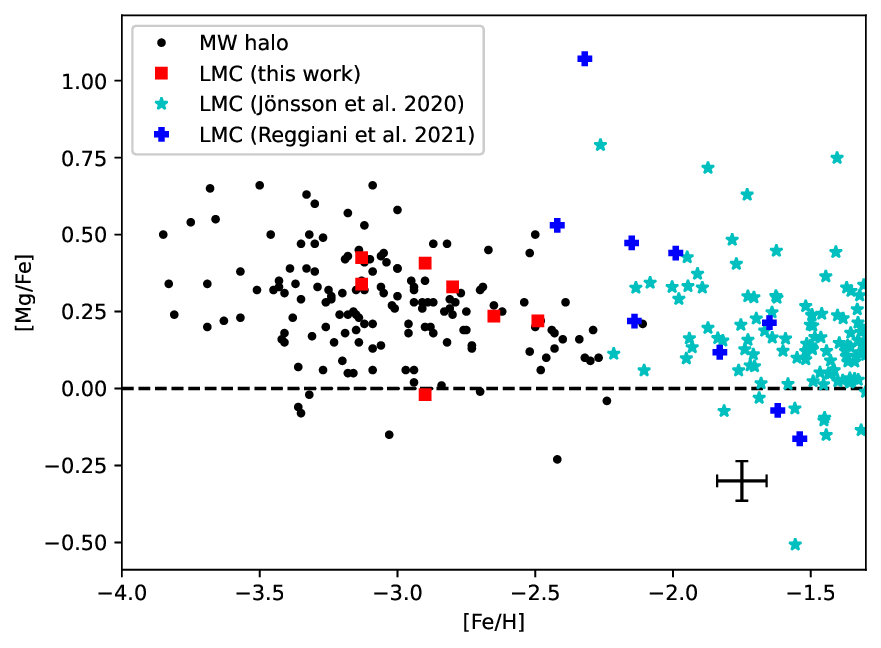}
    \caption{[X/Fe] vs [Fe/H] for elements C--Mg, for our Magellanic metal-poor sample (red triangles) compared to a Milky Way halo sample (black dots) from \citet{Yong2021}, LMC giants from \citet{Reggiani2021} (blue plus signs) and \citet{Jonsson2020} (cyan stars). A representative error bar for our data is provided in the bottom right corner. Upper limits to abundances are indicated by arrows. For [C/Fe], both the observed and corrected abundances are shown. For [O/Fe], 4 out of 7 stars were affected by telluric contamination that did not allow estimating an abundance or a meaningful limit. The dotted lines in the [C/Fe] and [N/Fe] plots indicate the lower limits of the CEMP ([C/Fe] > 0.7) and NEMP ([N/Fe] > 0.5) stars respectively.}
    \label{fig:X_Fe}
\end{figure*}

\begin{figure*}
        \includegraphics[width=\columnwidth]{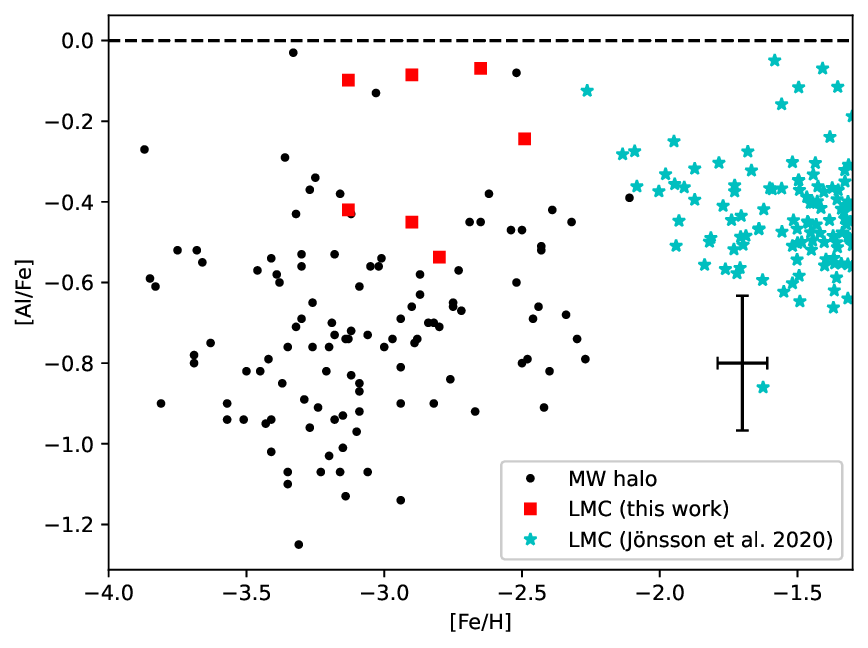}
        \includegraphics[width=\columnwidth]{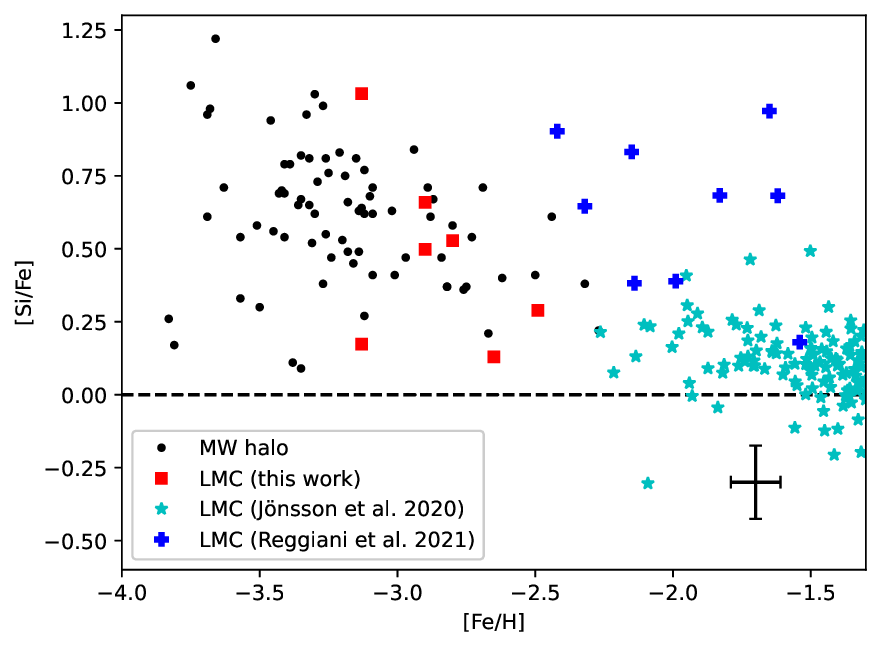}
        \includegraphics[width=\columnwidth]{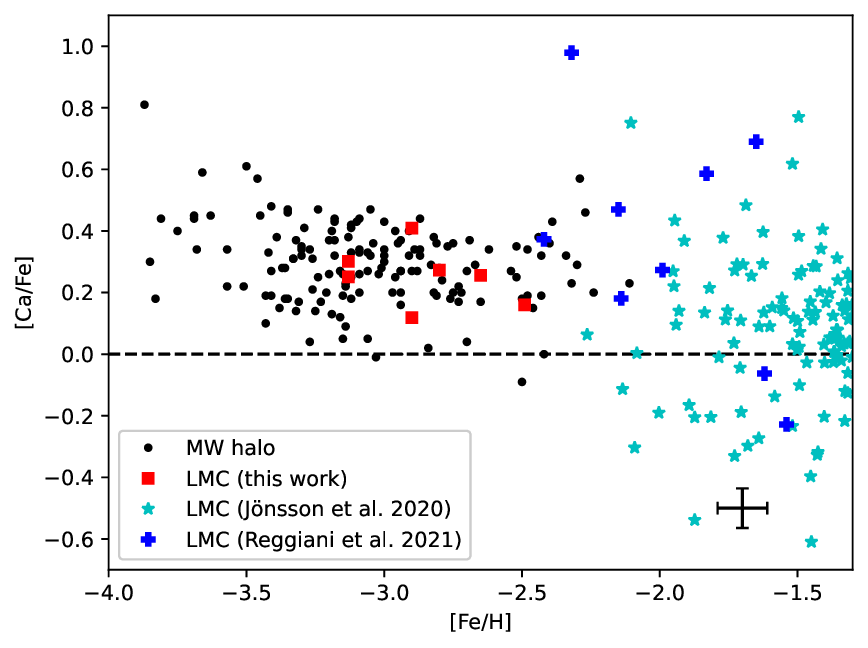}
        \includegraphics[width=\columnwidth]{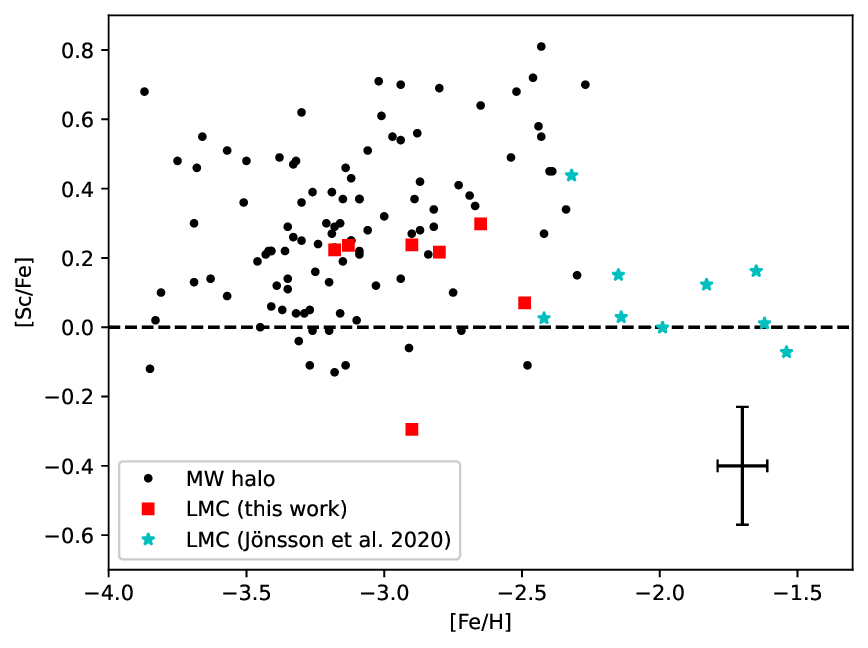}
        \includegraphics[width=\columnwidth]{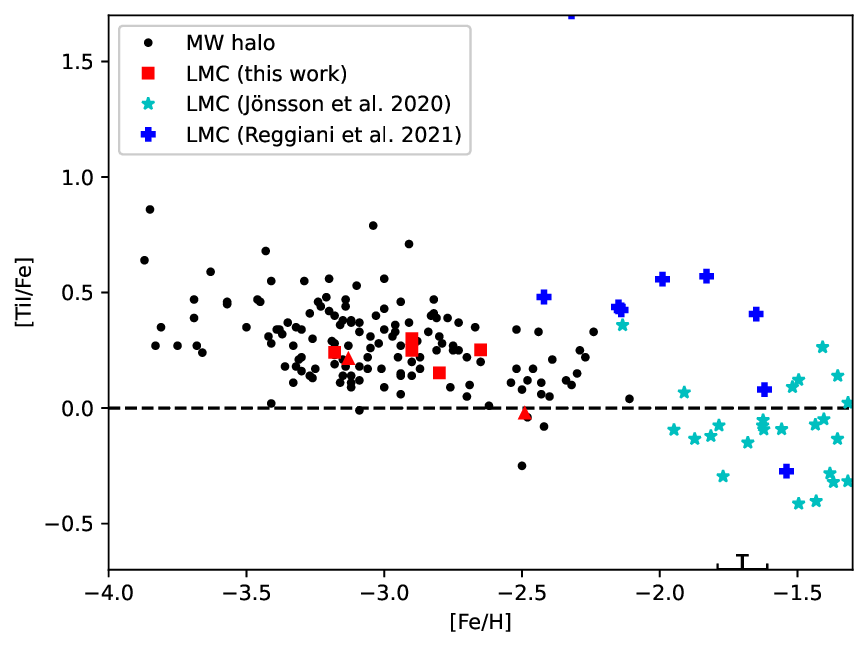}
        \includegraphics[width=\columnwidth]{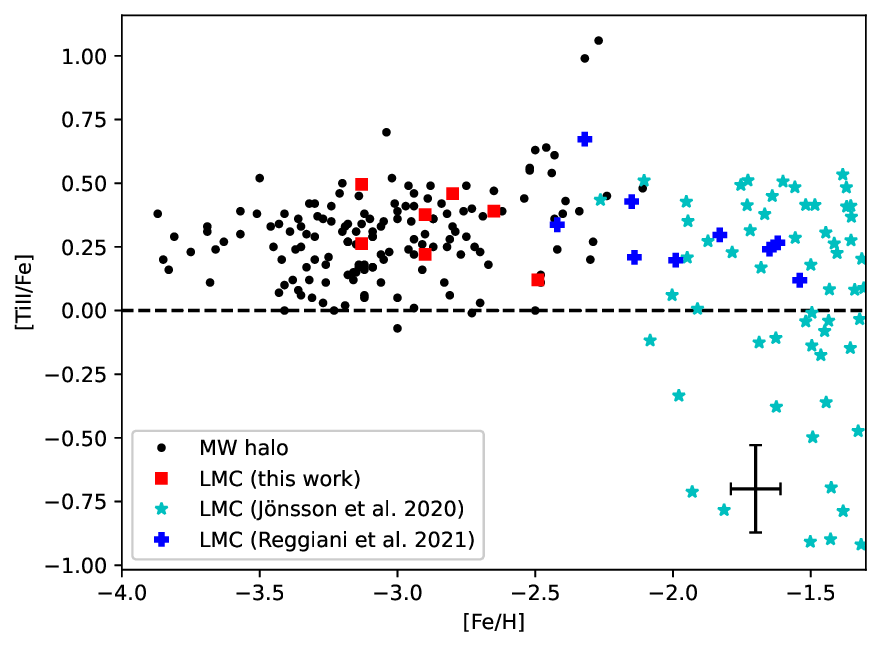}
    \caption{Same as Fig. \ref{fig:X_Fe} but for elements Al--Ti.}
    \label{fig:X_Fe_2}
\end{figure*}

\begin{figure*}
        \includegraphics[width=\columnwidth]{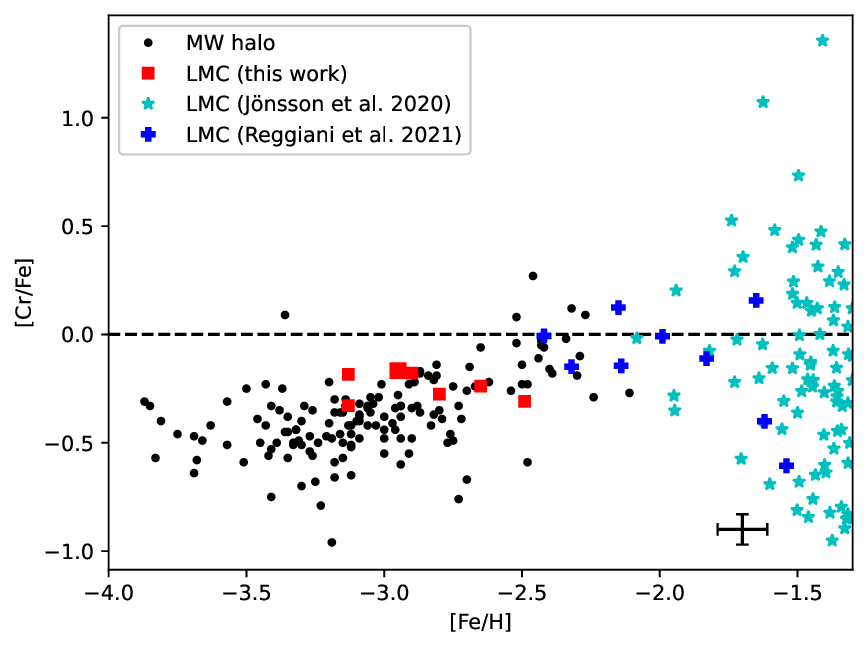}
        \includegraphics[width=\columnwidth]{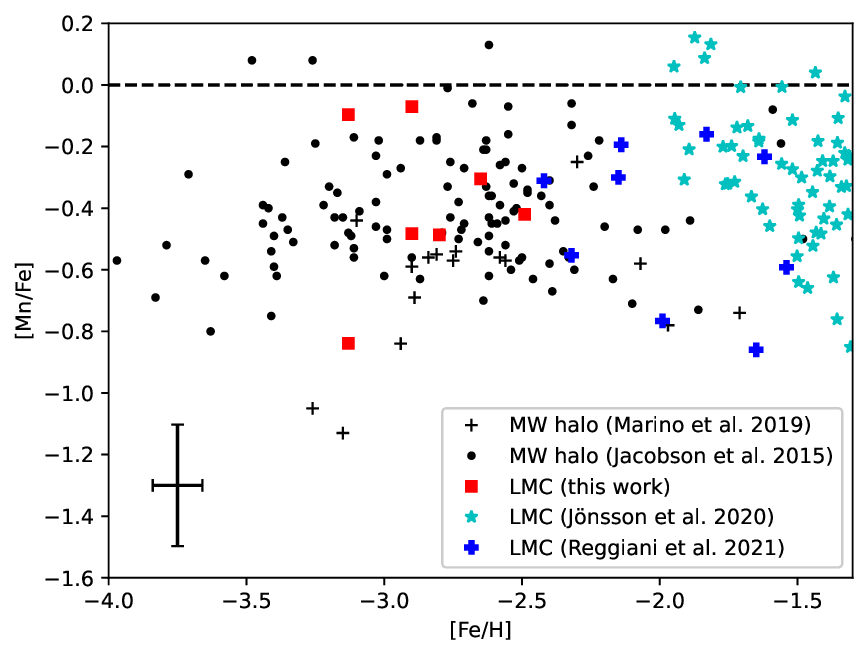}
        \includegraphics[width=\columnwidth]{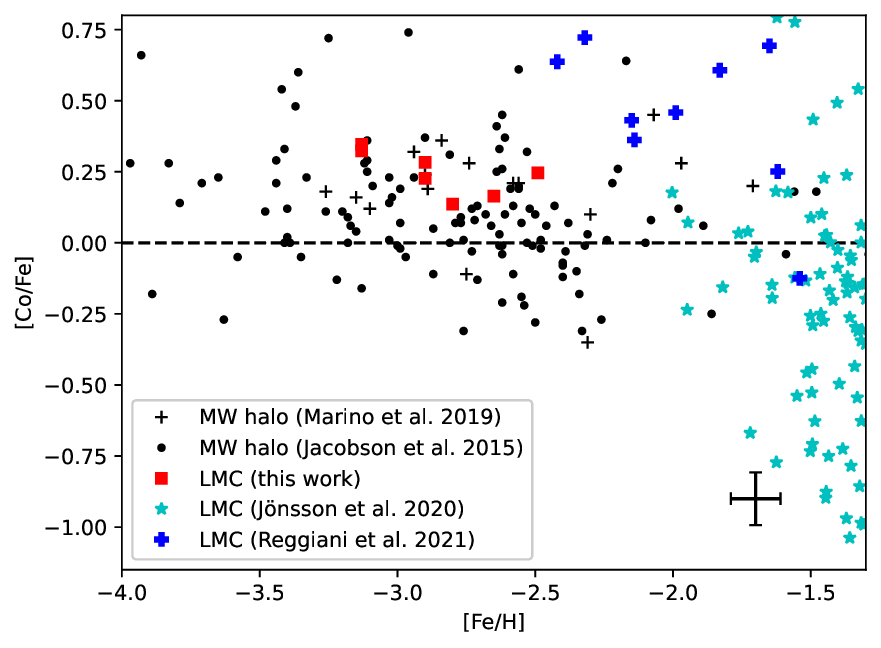}
        \includegraphics[width=\columnwidth]{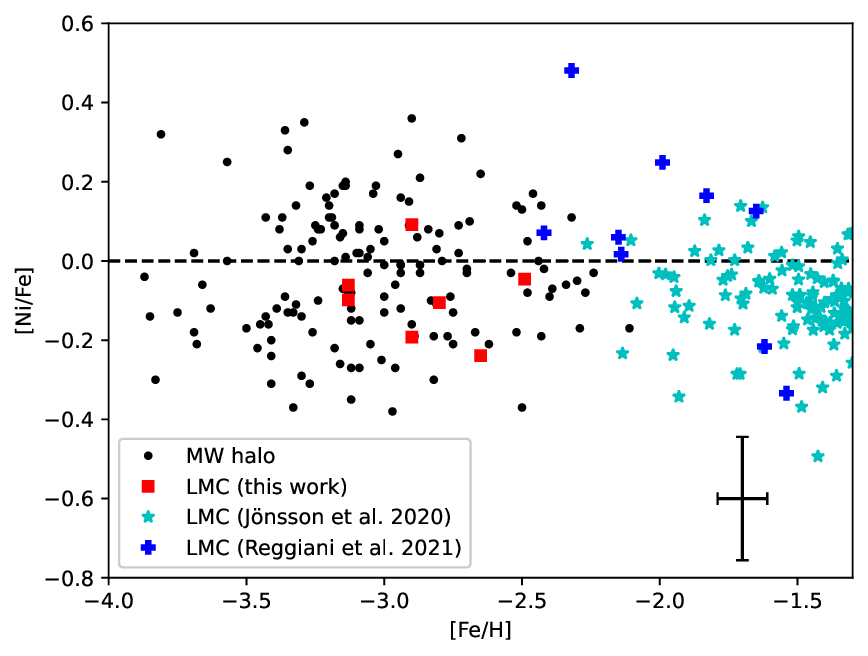}
        \includegraphics[width=\columnwidth]{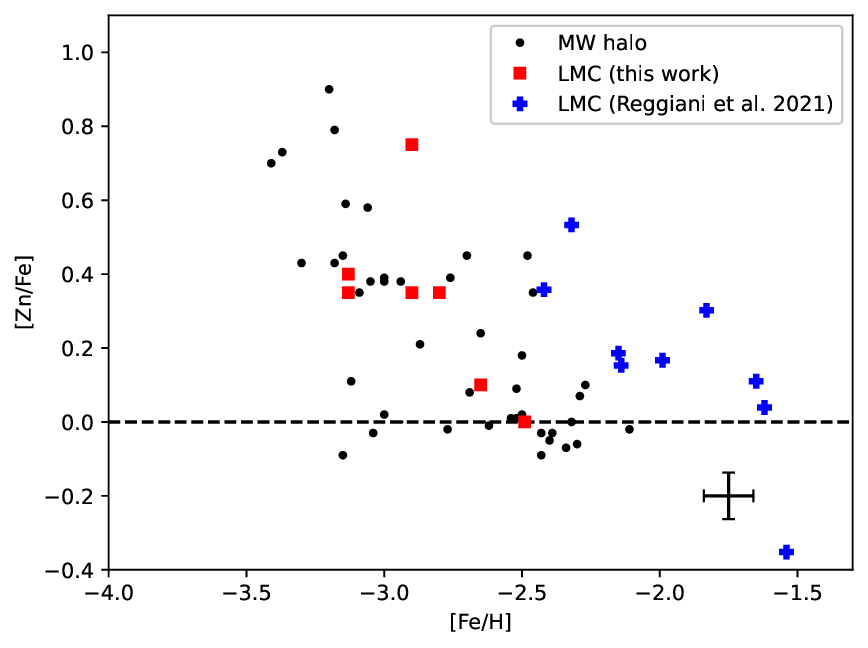}
        \includegraphics[width=\columnwidth]{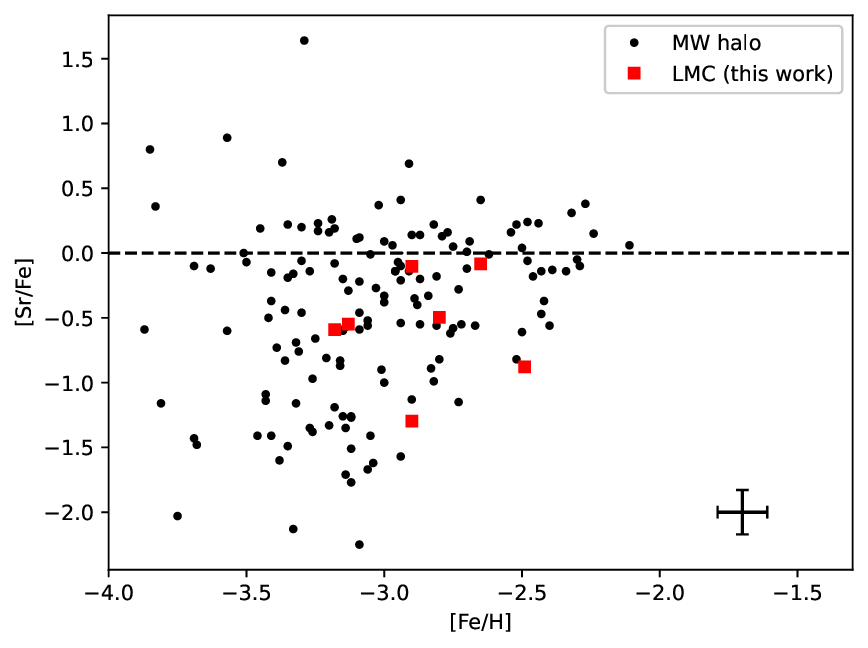}
    \caption{Same as Fig. \ref{fig:X_Fe} but for elements Cr--Sr. For Mn and Co, results from \citet{Jacobson2015} and \citet{Marino2019} were used instead of \citet{Yong2021} as mentioned in Section~\ref{4.2.3}.}
    \label{fig:X_Fe_3}
\end{figure*}

\begin{figure}
        \includegraphics[width=\columnwidth]{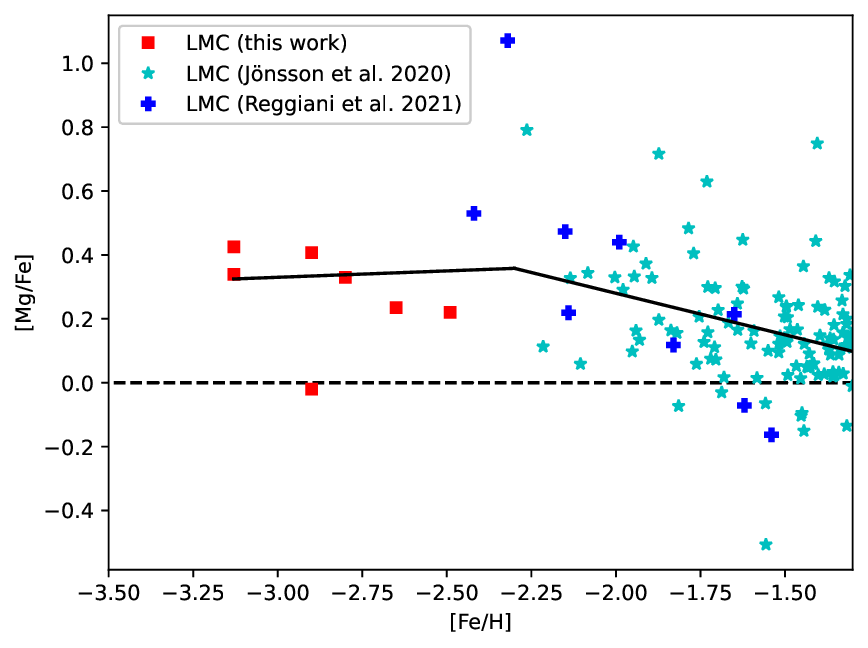}
    \caption{$\rm [Mg/Fe]$ for LMC stars together with data from \citet{Reggiani2021} and \citet{Jonsson2020}. A piecewise linear fit is overlaid, indicating an $\alpha$-knee at $\rm [Fe/H] = -2.3 \pm 0.4$.}
    \label{fig:alpha_knee}
\end{figure}

Based on the $\rm{[Mg/Fe]}$ plot in Fig.~\ref{fig:X_Fe}, our results seem to provide a low-metallicity anchor to the $\alpha$-knee\footnote{The $\alpha$-knee is defined by the value of [Fe/H] where the [Mg/Fe] or [$\alpha$/Fe] values starts to decrease from a near constant value at lower [Fe/H].}, which is caused by the occurrence of SNIa that produce large amounts of iron-peak elements relative to $\alpha$-elements, as compared to core-collapse supernovae. In a similar approach to \citet{Reggiani2021}, we used the \textsc{segmented} package in Python (\citealt{Muggeo2003}; \citealt{Pilgrim2021}) to quantitatively determine the $\alpha$-knee in the LMC. This involved fitting a segmented piece-wise linear model to the [Fe/H]--[Mg/Fe] plot as shown in Fig.~\ref{fig:alpha_knee}. Our result indicates that the $\alpha$-knee location is found at around $\rm{[Fe/H]} = -2.3 \pm 0.4$, and it is not inconsistent with that of \citet{Nidever2020}, which used the APOGEE data from \citet{Jonsson2020} to conclude that the $\alpha$-knee is constrained to about $\rm{[Fe/H]} \leq -2.2$. However, we note that our $\alpha$-knee determination has a large degree of uncertainty ($0.4$\,dex), and other $\alpha$-elements, such as calcium and titanium, do not show a clear $\alpha$-knee. Thus, more abundance measurements across the potential $\alpha$-knee location are needed to improve the precision of the determination.

\subsubsection{Iron-peak Elements}
\label{4.2.3}

The iron-peak elements include scandium, chromium, manganese, cobalt, nickel and zinc. Since the metallicities of our sample are likely located above the $\rm{[\alpha/Fe]}$ knee, the SNIa contribution to the iron-peak abundance ratios is very limited. Thus, the main contributor to the synthesis of these elements are core collapse (type II) supernovae.

For scandium, our results agree with that of \citet{Jonsson2020} at higher [Fe/H], but are slightly lower than that of the MW halo. One star, 500766372, has an unusually low [Sc/Fe] value ($\rm{[Sc/Fe]=-0.30 \pm 0.13}$). It also has high zinc, and very low strontium and barium abundances, and will be discussed further in Section \ref{sec:4.4.2}. After excluding the discrepant star, the average and scatter of the scandium abundances become $\rm{\langle[Sc/Fe]\rangle=0.08 \pm 0.21}$. There is little evidence to suggest that there is an intrinsic spread in [Sc/Fe] relative to the uncertainty of our measurements.

For cobalt and manganese, as our Milky Way comparison sample, we used results from \citet{Jacobson2015} and \citet{Marino2019} instead of \citet{Yong2021}. This is due to the fact that the [Co/Fe] and [Mn/Fe] results from \citet{Yong2021} appear inconsistent with other datasets, for reasons that are not entirely clear, but which may be related to the implementation of hyperfine structure corrections. No other element ratios are affected. 

In general, our iron-peak abundance measurements are consistent with those of the Milky Way halo. The typical measurement error and intrinsic spread in the abundance ratios relative to iron are on the order of 0.1--0.2 dex, which is also mostly consistent with the LMC determinations at higher [Fe/H]. One exception however, is zinc, with one star (500766372) having a high zinc abundance ($\rm{[Zn/Fe]=0.75 \pm 0.04}$) compared to the rest, similar to the highest MW values. For the other stars, 4 have $\rm{\langle[Zn/Fe]\rangle}\sim0.40$ while the remaining two most metal-rich stars have solar [Zn/Fe] comparable to the lowest [Zn/Fe] ratios in the MW halo. The other exception is scandium, where the star with high [Zn/Fe] also has an unusually low [Sc/Fe] value. In addition, the average abundances of Cr and Co are 0.1 dex higher than the MW average, but a little lower than Reggiani's typical abundances and in particular exhibit less scatter.

\subsubsection{Sr, Y, Zr and Ba}

At solar metallicities, heavy elements such as strontium, yttrium, zirconium and barium are produced by the slow neutron capture process (s-process) that occurs in asymptotic giant branch (AGB) stars, which are low to intermediate-mass stars in the late stages of their evolution (e.g. \citealt{Karakas2007}; \citealt{Kobayashi2020}). However, at low metallicities such as the ones found in our stars, it is not clear that there has been enough time to allow low to intermediate-mass AGB stars to evolve and distribute these elements into the LMC interstellar medium. 

For strontium,  our average abundance and standard deviation ($\rm{\langle[Sr/Fe]\rangle=0.25 \pm 0.43}$) are consistent with the MW halo, particularly in the existence of a large observed range of over 1 dex seen in the [Sr/Fe] measurements.

\begin{figure*}
        \includegraphics[width=\columnwidth]{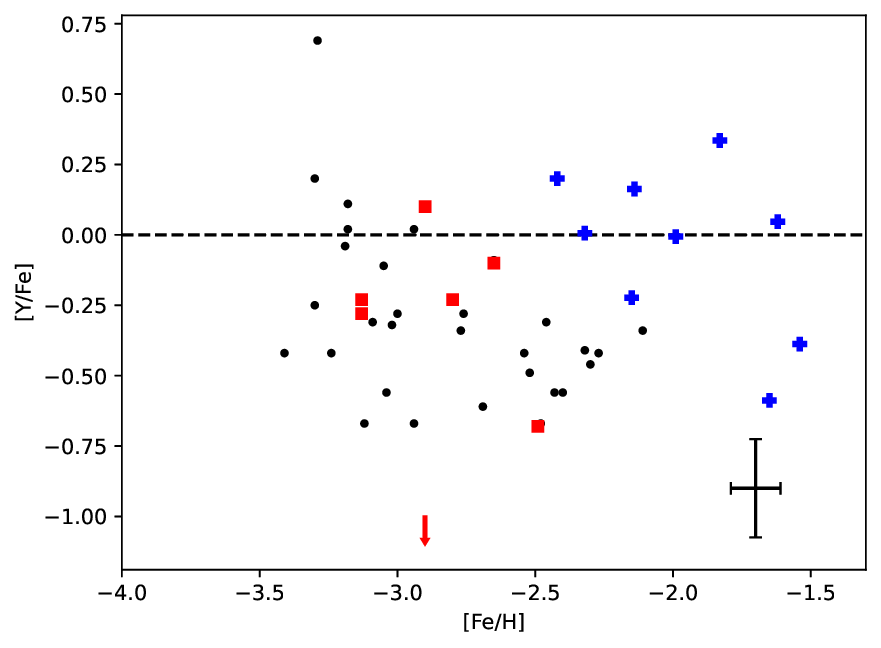}
        \includegraphics[width=\columnwidth]{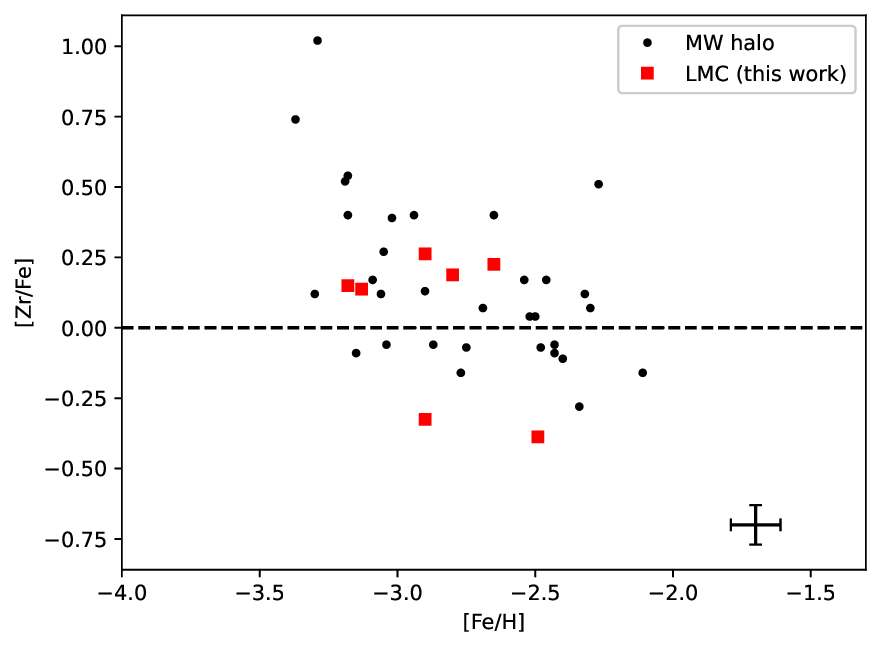}
        \includegraphics[width=\columnwidth]{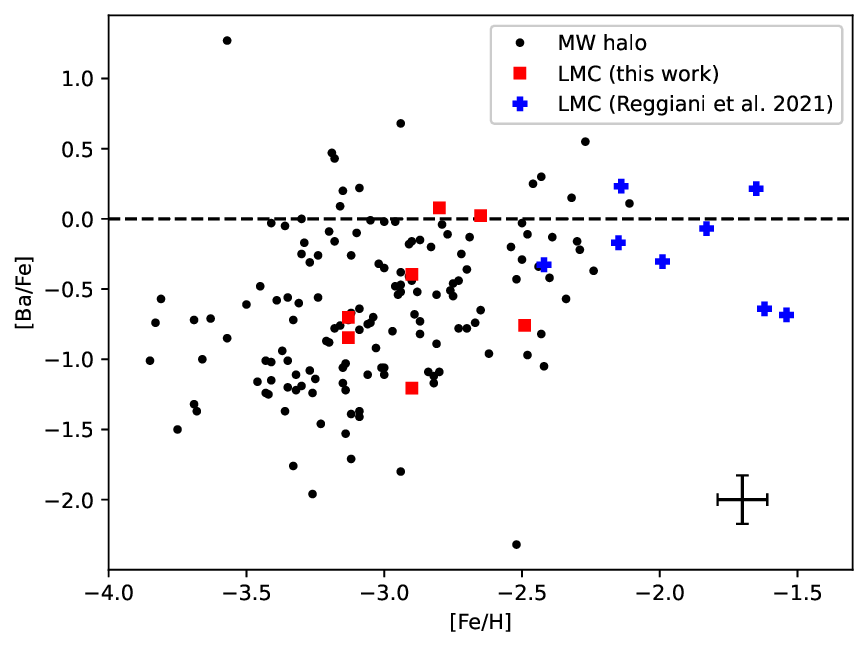}
        \includegraphics[width=\columnwidth]{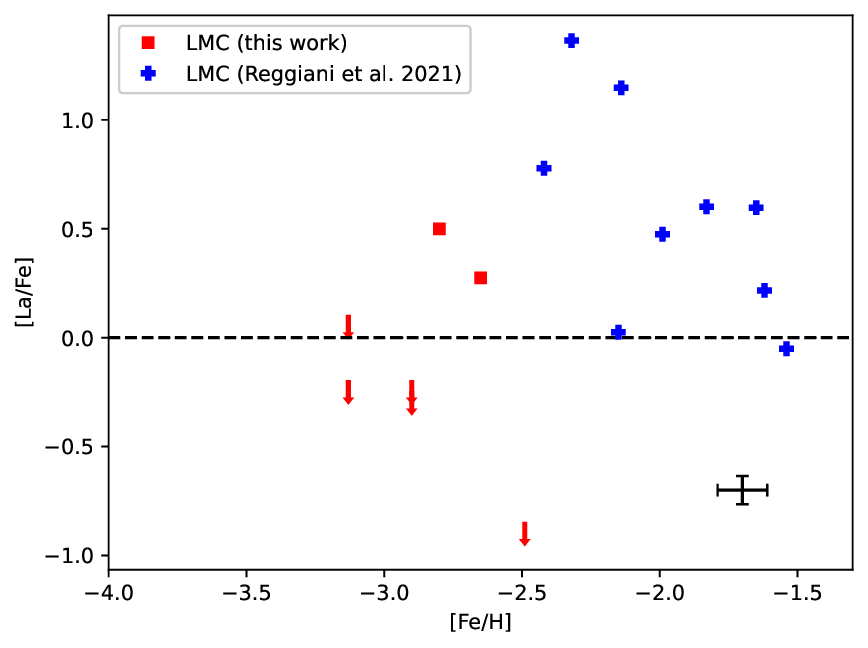}
        \includegraphics[width=\columnwidth]{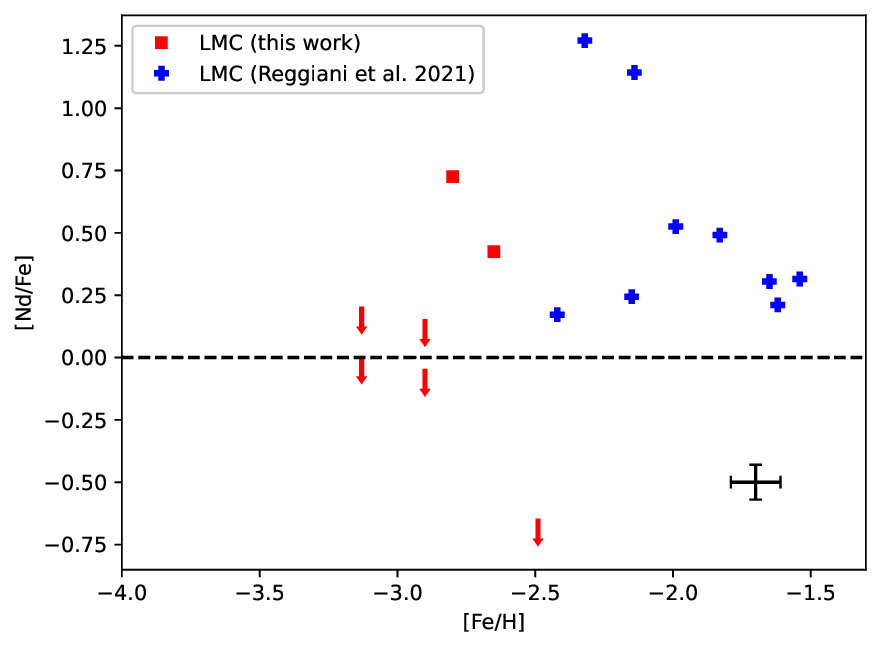}
        \includegraphics[width=\columnwidth]{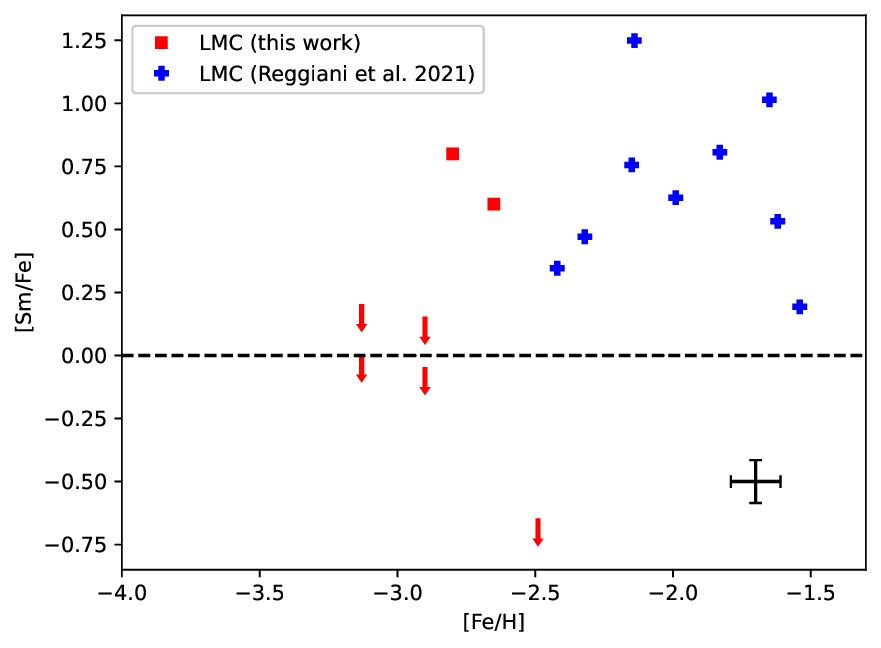}
    \caption{Same as Fig. \ref{fig:X_Fe} but for elements Y--Sm.}
    \label{fig:X_Fe_4}
\end{figure*}

For yttrium, our average abundance and standard deviation ($\rm{\langle[Y/Fe]\rangle=-0.24 \pm 0.26}$) are also consistent with MW halo and with the stars from \citet{Reggiani2021} that have higher [Fe/H]. There is, however, one star (500766372) in our sample with a very low abundance value $(\rm{[Y/Fe]} \leq -1.0)$, as shown by the upper limit in the plot shown in Fig.~\ref{fig:X_Fe_4}. It is important to point out however, that the measurements in our comparison MW sample are lacking due to limited S/N and wavelength coverage in \citet{Yong2021}. The true range of abundances in the MW may therefore be underestimated for yttrium.

For zirconium, similar to the previous two elements, our average abundance and standard deviation ($\rm{\langle[Zr/Fe]\rangle=0.04 \pm 0.27}$) are consistent with the MW halo. There are also two stars (500287810 \& 500766372) with evidently low zirconium abundances ($\rm{[Zr/Fe]=-0.39 \pm 0.07}$; $\rm{[Zr/Fe]=-0.33 \pm 0.08}$) as shown in the plot. This again may reflect selection effects in the MW comparison sample, which may suffer from restrictions in S/N and wavelength coverage \citep{Yong2021}.

For barium, our average abundance and standard deviation ($\rm{\langle[Ba/Fe]\rangle=-0.54 \pm 0.47}$) are again consistent with the MW halo, and they cover a large range in [Ba/Fe] (exceeding 1 dex) as seen in Figure~\ref{fig:X_Fe_4}. While \citet{Reggiani2021} found a fairly flat variation with metallicity, our barium abundances decline toward the lowest metallicities, in good agreement with the MW.

\begin{figure}
	\includegraphics[width=\columnwidth]{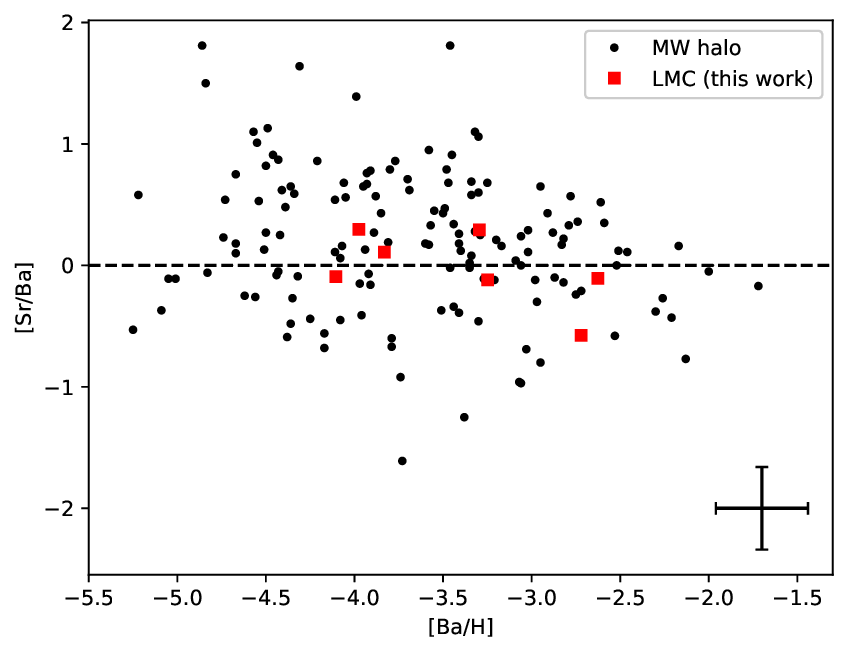}
    \caption{[Sr/Ba] vs [Ba/H] ratios for all 7 stars in our sample. The Milky Way halo results from \citet{Yong2021} have been included for comparison.}
    \label{fig:Sr_Ba}
\end{figure}

In Figure~\ref{fig:Sr_Ba}, we consider the [Sr/Ba] vs [Ba/H] ratios. \citet{Yong2013} found that at $\rm{[Ba/H]} < -2.5$, the intrinsic spead in [Sr/Ba] increases as [Ba/H] decreases. In contrast, our results do not exhibit an increase in intrinsic spread towards lower [Ba/H] values. Instead, the [Sr/Ba] ratios in our sample show a uniformly smaller spread than the halo at similar [Ba/H]. The low [Ba/H] and [Sr/H] values nevertheless indicate that there has not been sufficient time for AGB enrichment to have affected our stars.  Furthermore, the lack of high [Sr/Ba] values also suggests that any contribution from spinstars, which are thought to be a possible source of s-process enrichment at low metallicities \citep{Cescutti2014}, is not significant. Further details on this topic will be discussed in Section~\ref{sec:4.4.1}.

In general, the star-to-star scatter is larger than the measurement error for all four of these elements. This implies that the production is decoupled from that of the iron-peak, either because a second site is responsible for this production, or that the production depends on an parameter such as rotation (e.g. \citealt{Frischknecht2016}).

\subsubsection{Abundance summary}

Figure~\ref{fig:all_elements} summarises the abundance results for our LMC stars into a single plot. The MW halo results from the literature (\citealt{Cayrel2004}; \citealt{Jacobson2015}; \citealt{Marino2019}; \citealt{Yong2021}), which are averaged for $-3.13 \leq \rm{[Fe/H]} \leq -2.49$, are also shown for comparison. The red stars represent abundance measurements that differ by more
than 0.5 dex relative to the mean MW values (black dots and lines). Overall, our LMC results are consistent with that of the MW halo for most of the elements measured. We note however, that each star is strongly discrepant, by at least 0.5 dex compared to the MW, in at least one element.

\begin{table}
	\centering
	\caption{Abundance results, showing the sample size, mean, standard deviation and average measurement uncertainty} of our various abundance measurements, for stars where a given element was detected. Average abundances gathered for the Milky Way from the literature for $-3.13 \leq \rm{[Fe/H]} \leq -2.49$ \citep{Cayrel2004,Jacobson2015,Marino2019,Yong2021} have been included for comparison.
	\label{tab:elements_table}
	\begin{tabular}{lccccccc} % xxx columns, alignment for each
		\hline
		Element & $\rm{N}$ & $\rm{\mu}$ &  $\rm{\sigma_{dispersion}}$ & $\rm{\sigma_{err}}$ & $\rm{N_{MW}}$ & $\rm{\mu_{MW}}$ & $\rm{\sigma_{MW}}$\\
		\hline
            $\rm{[C/Fe]}$	&7	&0.16& 0.33	&0.20 &57	&0.33&0.3\\
		$\rm{[N/Fe]}$	&1	&1.70&0.00&	0.11 &13	&0.53&0.74\\
            $\rm{[O/Fe]}$	&3	&1.00&0.13&0.18 &14	&0.69&0.17\\
            $\rm{[Na/Fe]}$	&7	&0.27&0.45	&0.11 &63	&0.28&0.31\\
            $\rm{[Mg/Fe]}$	&7	&0.28&0.15&0.06&67	&0.27&0.14\\
            $\rm{[Al/Fe]}$	&7	&$-0.27$&0.20&0.17&44	&$-0.68$	&0.21\\
            $\rm{[Si/Fe]}$	&7	&0.47&0.31	&0.13& 26	&0.53&0.16\\
            $\rm{[Ca/Fe]}$	&7	&0.25&0.09	&0.06&69	&0.28&0.11\\
            $\rm{[Sc/Fe]}$	&7	&0.14&0.20 &0.17& 37	&0.36&0.20\\
            $\rm{[\ion{Ti}I/Fe]}$&7	&0.20&0.11	&0.06& 66	&0.26&0.16\\
            $\rm{[\ion{Ti}{II}/Fe]}$&7	&0.33&0.14 &0.17& 69	&0.30&0.16\\
            $\rm{[Cr/Fe]}$	&7	&$-0.24$&0.07 &0.07& 69	&$-0.35$&0.15\\
            $\rm{[Mn/Fe]}$	&7	&$-0.39$&0.26 &0.20& 70	&$-0.40$&0.18\\
            $\rm{[\ion{Fe}I/H]}$	&7	&$-2.86$&0.24 &0.09& 69	&$-2.89$&0.17\\
            $\rm{[\ion{Fe}{II}/H]}$&7	&$-2.83$&0.20	&0.09& 68	&$-2.86$&0.18\\
            $\rm{[Co/Fe]}$	&7	&0.25&0.08	&0.09&71	&$0.14$&0.26\\
            $\rm{[Ni/Fe]}$	&7	&$-0.09$&0.11	&0.16&68	&$-0.04$&0.16\\
            $\rm{[Zn/Fe]}$	&7	&0.33&0.24	&0.06&21	&0.20&0.18\\
            $\rm{[Sr/Fe]}$	&7	&$-0.57$&0.43	&0.17&68	&$-0.43$&0.60\\
            $\rm{[Y/Fe]}$	&6	&$-0.24$&0.26	&0.19&14	&$-0.37$&0.21\\
            $\rm{[Zr/Fe]}$	&7	&0.04&0.27	&0.07&15	&0.12&0.17\\
            $\rm{[Ba/Fe]}$	&7	&$-0.54$&0.47	&0.17&67	&$-0.62$&0.50\\
            $\rm{[La/Fe]}$	&2	&0.39&0.16	&0.07&0	&-&-\\
            $\rm{[Nd/Fe]}$	&2	&0.58&0.21	&0.07&0	&-&-\\
            $\rm{[Sm/Fe]}$	&2	&0.70&0.14	&0.09&0	&-&-\\
            $\rm{[Eu/Fe]}$	&2	&0.91&0.16	&0.07&15	&0.31&0.41\\
            $\rm{[Dy/Fe]}$	&2	&0.93&0.11	&0.08&0	&-&-\\
            \hline
	\end{tabular}
\end{table}

\begin{figure}
        \includegraphics[width=\columnwidth]{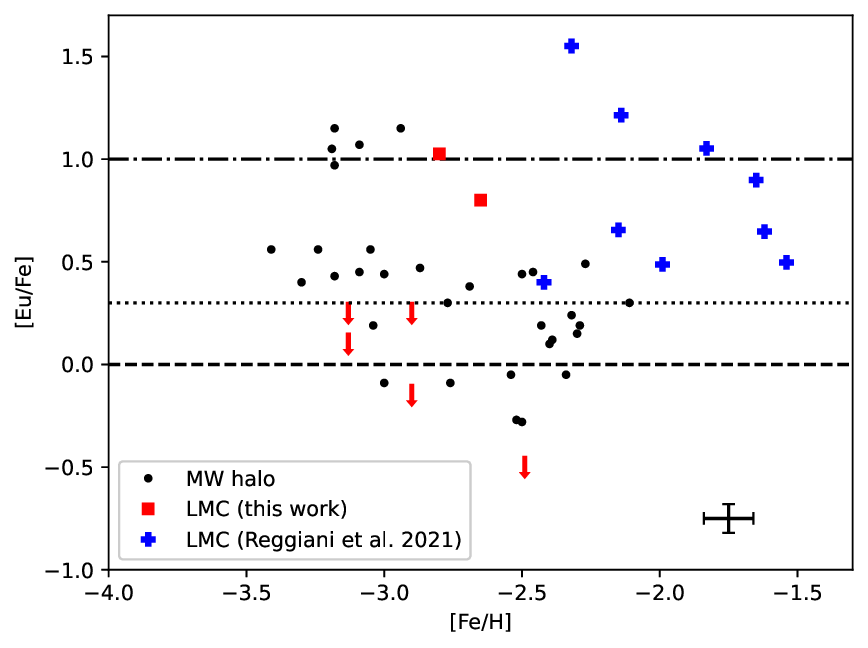}
        \includegraphics[width=\columnwidth]{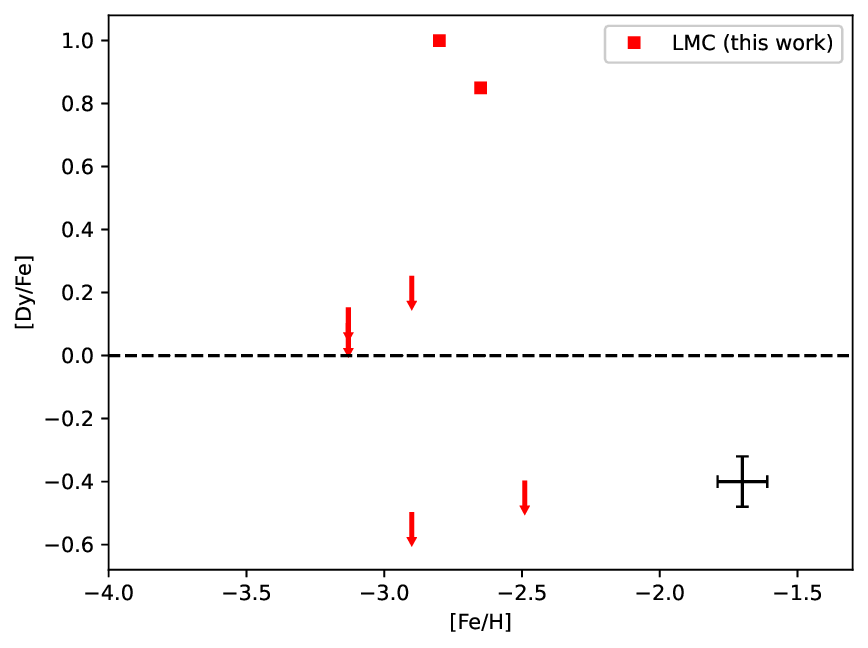}
    \caption{Same as Fig. \ref{fig:X_Fe} but for Eu and Dy. For the Eu plot, the dotted and dash-dot lines represent the lower limit for r-I ([Eu/Fe] > 0.3) and r-II ([Eu/Fe] > 1.0) stars respectively.}
    \label{fig:X_Fe_5}
\end{figure}

\subsection{r-process elements} 
\label{sec:4.3}

\begin{figure}
	\includegraphics[width=\columnwidth]{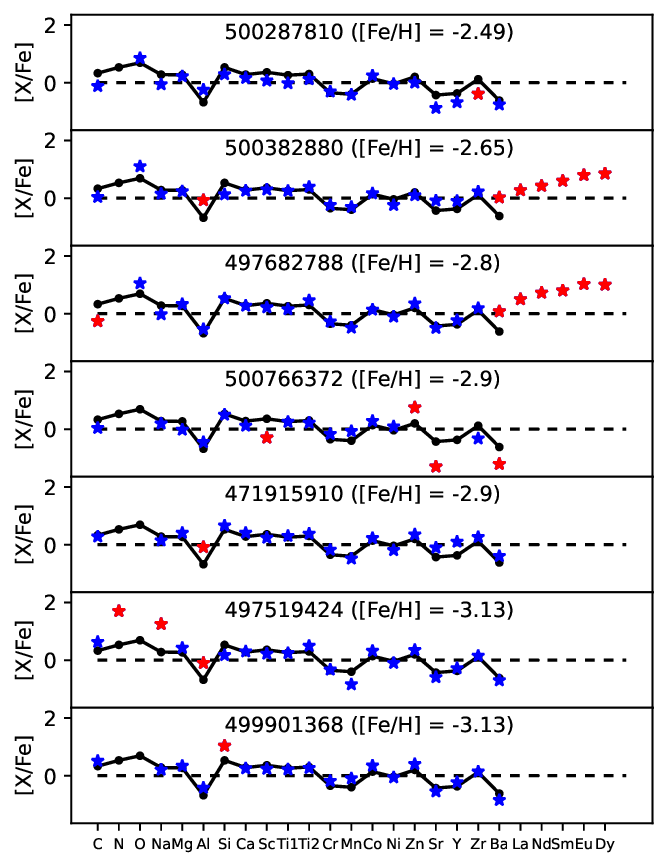}
    \caption{Overview of the abundance measurements for individual stars (500287810, 500382880, 497682788, 500766372, 471915910, 497519424, 499901368) with literature values.
    The stars represent our abundance measurements, and the black dots represent the mean MW values from the literature (\citealt{Cayrel2004}; \citealt{Jacobson2015}; \citealt{Marino2019}; \citealt{Yong2021}). The star colours indicate if the abundance is within 0.5 dex of the MW average (blue) or not (red).}
    \label{fig:all_elements}
\end{figure}

\begin{figure}
	\includegraphics[width=\columnwidth]{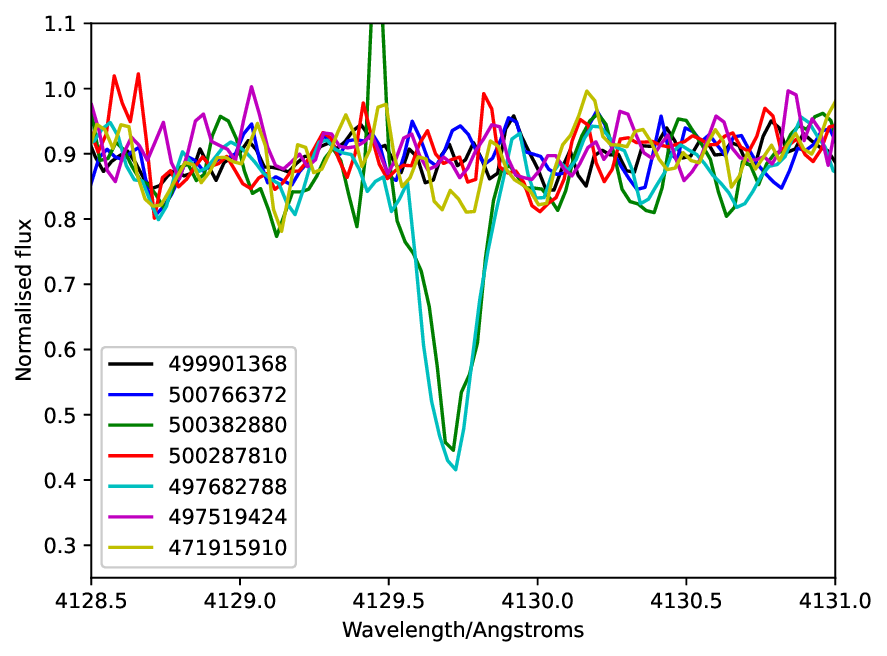}
    \caption{The [\ion {Eu} {II}] line at 4129 \AA \hspace{0.03cm} for all 7 stars in our sample. While the line cannot be detected in 5 out of 7 stars, the two stars 497682788 and 500382880 are clearly over-abundant. For these stars, we estimate their Eu abundances to be $\rm{[Eu/Fe]} = 1.03 \pm 0.06$ and $\rm{[Eu/Fe]} = 0.80 \pm 0.08$ respectively.}
    \label{fig:Eu_4129}
\end{figure}

As shown in the $\rm{[Eu/Fe]}$ plot in Fig.~\ref{fig:X_Fe_5} and in the spectra in Fig.\ \ref{fig:Eu_4129}, two of our stars are r-process enhanced. Using the customary classifications for r-process enhanced stars ($\rm[Eu/Fe] > 1.0$ and $\rm[Ba/Eu] < 0.0$; \citealt{Barklem2005}), 497682788 is an r-II star. The other one is 500382880, an r-I star ($0.3 < \rm[Eu/Fe] < 1.0$ and $\rm[Ba/Eu] < 0.0$). Since the constraint for pure r-process production of Ba and Eu is thought to be $\rm{[Ba/Eu]}=-0.78\pm0.06$ \citep{Mashonkina2014}, our stars' relatively low abundance ratios ($\rm \langle[Ba/Eu]\rangle \approx -0.9$) most likely indicates a pure r-process enrichment. For the other 5 stars, only upper limits could be measured for the [Eu/Fe] ratios. Their values clearly show that they are not significantly r-process enhanced. This result is confirmed by comparing the abundance ratios for the other r-process elements (La, Nd, Sm, Dy) in Fig.~\ref{fig:all_elements}; they all paint the same picture in showing that 2 out of the 7 stars are r-process enhanced, and the other 5 are not. 

We also compare the r-I and r-II star occurrences in the LMC and the Milky Way halo. For our sample, the results reveal that the occurrence rates of these stars are statistically indistinguishable in the two environments. Specifically, the LMC sample exhibited occurrence rates of 1/7 ($14_{-9}^{+16}$\%) for r-I stars, 1/7 ($14_{-9}^{+16}$\%) for r-II stars and 2/7 ($29_{-14}^{+17}$\%) for both r-I and r-II stars, while the Milky Way halo sample \citep{Yong2021} exhibited occurrence rates of 16/150 ($11_{-2}^{+3}$\%) for r-I stars, 4/150 ($3_{-1}^{+2}$\%) for r-II stars and 20/150 ($13_{-3}^{+3}$\%) for both r-I and r-II stars. The statistical likelihood analysis was performed following the approach of \citet{Reggiani2021}. However, if we factor in the 4 very metal-poor stars from \citet{Reggiani2021}, the occurrence rates for all the very and extremely metal-poor stars in the combined LMC sample (i.e., stars with $-3.13 \leq \FeH \leq -2.14$) increases to 3/11 ($27_{-11}^{+14}$\%) for r-I stars, 3/11 ($27_{-11}^{+14}$\%) for r-II stars and 6/11 ($55_{-14}^{+13}$\%) for both r-I and r-II stars. Our findings indicate that the r-II fraction deviates by more than 2 sigma from the Milky Way value, considering a similar [Fe/H] range as the sample in \citet{Yong2013}. To confirm or disprove the potential surplus of r-II stars relative to the Milky Way, a more extensive collection of LMC stars with [Eu/Fe] measurements at $\rm [Fe/H] < -2$ is essential. These findings provide valuable insights into the similarities and potential differences of stellar populations in the LMC and the Milky Way halo, which becomes more clear at the lowest metallicities.

\begin{figure}
	\includegraphics[width=\columnwidth]{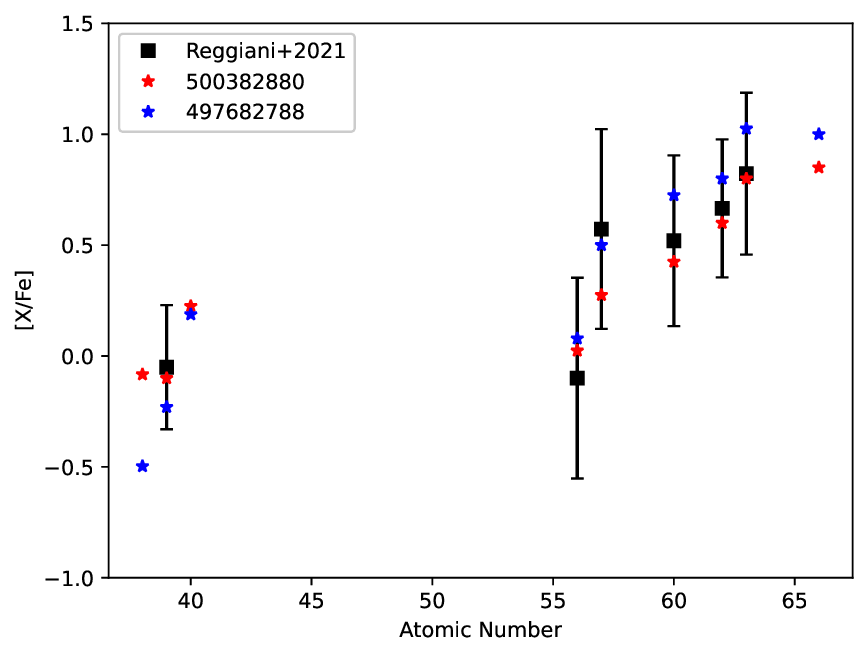}
    \caption{[X/Fe] ratios of heavy elements Sr--Dy for our 2 r-process enhanced stars, compared to the mean and scatter of the sample from \citet{Reggiani2021}. No rescaling has been applied to the abundances. }
    \label{fig:Reggiani_comparison}
\end{figure}

\begin{figure}
	\includegraphics[width=\columnwidth]{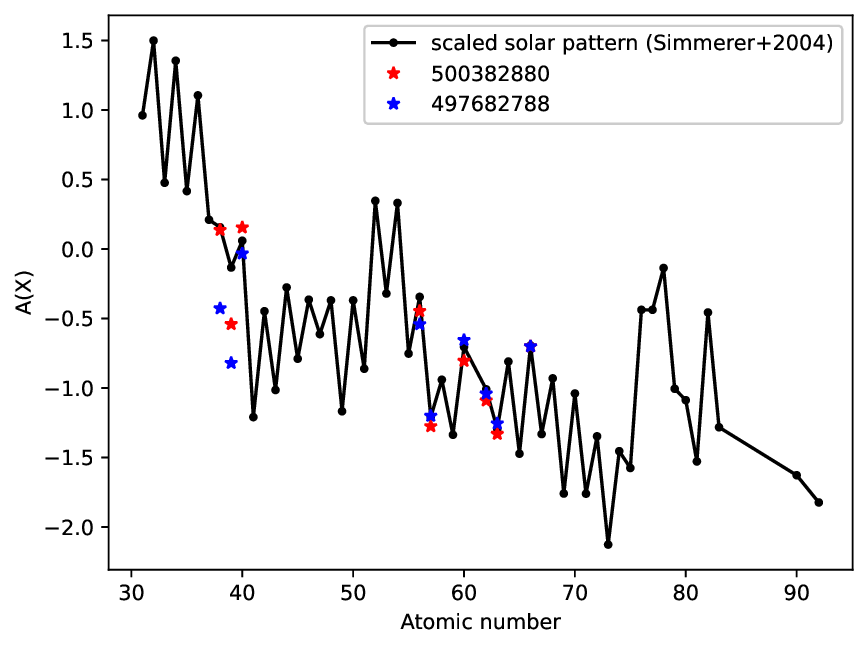}
    \caption{Heavy element abundances of Sr--Dy for our 2 r-process enhanced stars (500382880 \& 497682788), compared with the scaled solar r-process pattern from \citet{Simmerer2004}. The solar pattern has been normalised to the mean of the two A(Eu) values. Our abundances agree well with the solar scaled pattern for heavy elements $Z=56$--66 as well as $Z=40$, but less so for the light elements $Z=38$--39}
    \label{fig:test_absolute}
\end{figure}

For the 2 stars in our sample that are r-process enhanced, their r-process element abundance ratios relative to iron are consistent with those of \citet{Reggiani2021} as shown in Fig.~\ref{fig:Reggiani_comparison}. In addition, their abundances also seem to follow the solar r-process pattern \citep[e.g.][]{Simmerer2004}, as shown in Fig.~\ref{fig:test_absolute}. Our result is consistent with that of the literature, where there is a good agreement with scaled solar r-process abundances for heavier elements $(Z \geq 56)$ in other r-process rich stars \citep{Ernandes2023}. We also note that just like in the literature, our values do not agree as well with the solar abundance ratios for the lighter neutron-capture elements Sr and Y.

The heavy element abundances reported by \citet{Reggiani2021} exhibit a significant degree of scatter among their sample. While most of this can be attributed to the varying levels of r-process enhancement, it is worth noting that even after normalising by [Eu/Fe] or the average [X/Fe] values for heavy elements, a substantial scatter still remains. To better visualise the comparison, we opted to calculate a straight average of their abundances for each element without any renormalisation.

For our two r-process enhanced stars, the [X/Fe] abundances agree well with the average abundances from \citet{Reggiani2021} as shown in Fig.~\ref{fig:Reggiani_comparison}, indicating consistency in the r-process for all these r-process enhanced stars. In addition, their abundances of the two r-process enhanced stars also seem to follow the solar r-process pattern \citep[e.g.][]{Simmerer2004}, as shown in Fig.~\ref{fig:test_absolute}. Our result is consistent with that of the literature, where there is a good agreement with scaled solar r-process abundances for heavier elements $(Z \geq 56)$ in other r-process rich stars \citep{Ernandes2023}. We also note that just like in the literature, our values do not agree as well with the solar abundance ratios for the lighter neutron-capture elements Sr and Y.

\subsubsection{Implication of the r-process abundances}
The detection of low levels of barium ($\rm{[Ba/Fe]}\sim -0.8$) and strontium ($\rm{[Sr/Fe]}\sim -0.7$) in our two most metal-poor stars suggests the presence of some base level of r-process enrichment at the lowest metallicities, most likely originating from some form of core-collapse supernovae \citep[e.g.][]{Ji2019}.

Additionally, the lack of r-process enhancements in those stars may indicate a longer timescale for the site of the r-process nucleosynthesis. There are a number of possible r-process sites, including magneto-rotational supernovae, but the major source is likely to be binary neutron star mergers, although this topic continues to be debated within the community \citep{Kobayashi2023}. Binary neutron star mergers are known to occur on extended timescales, ranging from $\sim30$ Myr to several Gyr \citep{Belczynski2018, Neijssel2019}, while magneto-rotational supernovae are related directly to massive star evolutionary timescales, on the order of 10--20 Myr \citep{Ji2019}. Therefore, based on the absence of r-process enhancements in the most metal-poor stars ($\FeH < -2.8$) in our sample, it is plausible that magneto-rotational supernovae might have made only a minor contribution in the LMC. 

Furthermore, chemical evolution models have noted that the MW reached $\FeH \sim -3.5$ after about 60 Myr \citep{Kobayashi2020}. \citet{Nidever2020} found in their model for the Large Magellanic Cloud (LMC) that chemical evolution had advanced no further than $\FeH < -2.2$ after about 100 Myr. Thus, based on the $\FeH = -2.8$ cutoff for r-process enrichment in our results, and taking into account the chemical evolution models, we postulate a minimum timescale of $\sim100$ Myr for the neutron star binary merger process to generate substantial r-process enhancements in the LMC. This finding is in line with the conclusions drawn by \citet{Reggiani2021}, whose study focused on stars that are comparatively more metal-rich than those in our sample, and who found that all the stars in their sample exhibited r-process enhancement.

Interestingly, the findings from \citet{Ji2016} regarding the UFD Reticulum II (Ret II) are also consistent with our results. Just like our sample, the two most metal-poor stars in Ret II with $\FeH < -3$ are not r-process enhanced in the way the more metal-rich Ret~II stars are \citep{Ji2016}. While we cannot rule out the presence of inhomogeneous metal mixing in such environments, it is plausible that both the LMC and Ret II contain stars at low $\FeH$ that formed before major r-process enrichment events, indicating that significant r-process enrichment did not occur until after a degree of enrichment in the $\alpha$ and iron-peak elements.

However, the study of \citet{Skuladottir2019} shows continuous r-process enrichment in the Sculptor dwarf galaxy, indicating chemical evolution without a significant time delay relative to core-collapse supernovae. This disparity with our findings highlights the complexity in terms of comprehending r-process enrichment timescales. Additional europium abundance measurements for dwarf galaxies at lower metallicities could prove instrumental in addressing this issue.

\subsection{Individual stars}
\label{sec:4.4}

\subsubsection{497519424: NEMP}
\label{sec:4.4.1}
In Section~\ref{sec:4.2.1}, we demonstrated that 49751942 is a NEMP star, with elevated levels of sodium and aluminum. Building upon this information, we propose that the  chemical abundance pattern observed in this star may be indicative of enrichment from a rapidly rotating progenitor.

Previous studies, such as those conducted by \citet{Meynet2002}, \citet{Hirschi2007}, and \citet{Meynet2010}, have revealed that rapid rotation in massive stars plays a pivotal role in facilitating the upward transport of processed materials from the star's inner regions to its surface. This leads to an enhancement of specific elements in the stellar atmosphere, notably nitrogen, sodium, and aluminium, which are derived from distinct processes, namely CNO processing, the Na-Ne cycle, and the Mg-Al cycle, respectively. A massive rapidly rotating star would have ultimately dispersed these elements into the interstellar medium via stellar winds and eventual supernovae, contributing to the chemical evolution of the galaxy.

Comparing 497519424 with other NEMP stars in the Milky Way, we see that out of the 11 such stars in \citet{Yong2021}, only two are also enhanced in both sodium and aluminium. This could indicate that the processes responsible for these stars do not significantly depend on the galactic environment, since it is seen in both the LMC and the Milky Way.

However, it is important to highlight that some models of rapidly rotating stars also typically yield higher quantities of s-process elements via the weak s-process (\citealt{Pignatari2010}; \citealt{Cescutti2014}; \citealt{Frischknecht2016}). This expected enhancement is not detected in 497519424. This could stem from the extremely metal-poor nature of the progenitor star where the comparative lack of iron-peak seeds from which to generate the s-process elements may have hindered their formation.

In a more recent study by \citet{Choplin2018} of massive stars with $\rm [Fe/H] = -1.8$, it was observed that non-rotating ($\rm{v/v_{crit}=0}$) and slow-rotating massive ($\rm{v/v_{crit}=0.4}$) stars exhibit significant [N/Fe] enhancements. However, this pattern is rarely observed in rapidly rotating massive stars ($\rm{v/v_{crit}=0.7}$), where only 1 out of 9 simulations (with different stellar masses) had enhanced [N/Fe]. Therefore, the high N, Na and Al in 497519424 suggest rotation in the progenitor star, but the lack of high Sr, Y and Zr indicates that there has not been substantial accompanying weak s-process enrichment \citep{Pignatari2010}.

%Furthermore, across their various models, [Ba/Fe] enhancements do not appear to be a common occurrence, irrespective of rotational velocity \citep{Choplin2018}. 

% Here are the numbers I extracted from Choplin+ 2018, for stars producing [N/Fe] > 1.5:
% Vcrit    N
%       0.00000         462
%      0.400000         723
%      0.700000         100
%
% mass/Msol         N
%      10          25
%      15          32
%      20          79
%      25         197
%      40         161
%      60         191
%      85         200
%     120         200
%     150         200
% 
% In fact, all the v/vcrit=0.7 models have 25 Msol. There are 100 of them because they use different mass cuts.

\subsubsection{500766372: Hypernova signature}
\label{sec:4.4.2}

The chemical abundance pattern observed in star 500766372 is characterised by low levels of scandium, yttrium, strontium, and barium, but a high zinc abundance.  We propose that star was enriched with the nucleosynthetic products from a hypernova \citep{Kobayashi2014}.

Hypernovae are highly energetic supernovae resulting from the core collapse of massive stars. Compared to regular core-collapse supernovae, their significantly higher explosion energies are believed to lead to the production and ejection of large amounts of heavy elements. Studies on supernova nucleosynthesis, such as \citet{Nomoto2013}, have shown that hypernovae are potential sites for the production of zinc, which can be efficiently synthesised in the outer layers of the exploding star and then expelled into the interstellar medium. 

As a comparison, the EMP star SMSS J200322.54-114203.3 from \citet{Yong2021b} contains high zinc but not low scandium. It also exhibits substantial enhancement in the r-process elements which are hypothesized to originate in a magneto-rotational hypernova \citep{Yong2021b}. As a result, the enhanced zinc abundance together with the low scandium, strontium and barium abundances in 500766372 suggests that this star is not the result of a magneto-rotational hypernova event.

\subsubsection{500287810: Spinstar signature}
\label{sec:4.4.3}

This star has the highest barium to europium ratio amongst our LMC sample $(\rm{[Ba/Eu]} > -0.3)$. It has fairly low abundances in all the neutron-capture elements ($\rm{[Eu/Fe]} < -0.45$; $\rm{[La/Fe]} < -0.85$) including those that can be made in the s-process ($\rm{[Sr/Fe]} = -0.88$; $\rm{[Ba/Fe]} = -0.76$). Thus, although it did not form sufficiently late for s-process enrichment by AGB stars to become important, based on the [Ba/Eu] limit, the low abundances for the neutron capture elements seem to originate from another type of neutron capture-process enrichment rather than an r-process one. One potential source is spinstars, which are thought to be a possible site for weak s-process enrichment at low metallicities (\citealt{Pignatari2010}; \citealt{Cescutti2014}; \citealt{Choplin2020}). We note that this process often produces low Sr and Ba abundances with a range of [Sr/Ba] ratios, including the solar [Sr/Ba] ratio observed in star 500287810.

A further important result for this star is its lack of r-process enrichment ($\rm{[Eu/Fe]} < -0.45$) despite being the most metal-rich star in our sample. This probably suggests that r-process enrichment that happened after $\rm{[Fe/H]} \sim -3$ was not spatially homogeneous across the entire LMC as it existed $\sim12$ Gyr ago.  Otherwise every star above $\rm{[Fe/H]} \sim -3$ would show some level of r-process enrichment as is seen in our two r-process enhanced stars and in those from the \citet{Reggiani2021} sample.

\section{Summary}

In this paper, we have presented abundance results, based on high-resolution spectra, for 7 metal-poor stars that are members of the Large Magellanic Cloud.  Our analysis confirms that all 7 stars, including two extremely metal-poor (EMP) stars with [Fe/H] $\leq$ --3, are the most metal-poor stars discovered thus far in the Magellanic Clouds. Although their abundances and abundance ratios generally resemble those found in the Milky Way halo, our investigation reveals several key differences. Out of the seven stars in our sample, three stars potentially exhibit enrichment by rotating massive stars or hypernovae. Together with the results from \citet{Reggiani2021}, the absence of r-process enhancement in stars with lower [Fe/H] values suggests a minimum delay timescale of $\sim$100 Myr for the neutron star binary merger process to generate substantial r-process enhancements in the LMC. Furthermore, the occurrence rates of r-I and r-II stars are statistically indistinguishable in the very and extremely metal-poor stars in our LMC sample and the Milky Way halo. However, the r-II star occurrence rate is potentially higher for the LMC when the stars from \citet{Reggiani2021} are added to the comparison sample. Ultimately, these results provide valuable insights into the earliest stages of star formation in the LMC and offer the first comprehensive information on extremely and very metal-poor stars in galaxies of this mass.

\section*{Acknowledgements}

We would like to express our warmest gratitude to Dr David Yong and Madeleine McKenzie for their generous guidance and support with the spectroscopic analysis methods. We also thank the referee for their comments that have led to improvements in the paper. This paper includes data gathered with the 8m VLT located at Cerro Paranal, Chile, and is based on observations collected at the European Southern Observatory under ESO programme 108.22D8. This research was supported by the ARC Centre of Excellence for All Sky Astrophysics in 3 Dimensions (ASTRO 3D), through project number CE170100013.

%%%%%%%%%%%%%%%%%%%%%%%%%%%%%%%%%%%%%%%%%%%%%%%%%%
\section*{Data Availability}

The data used in this study are available in the ESO archive (https://archive.eso.org/eso/eso\_archive\_main.html) under programme ID 108.22D8. Our coadded spectra are available upon request.



\bibliographystyle{mnras}
\bibliography{EMP} % if your bibtex file is called example.bib

\begin{thebibliography}{}
\makeatletter
\relax
\def\mn@urlcharsother{\let\do\@makeother \do\$\do\&\do\#\do\^\do\_\do\%\do\~}
\def\mn@doi{\begingroup\mn@urlcharsother \@ifnextchar [ {\mn@doi@}
  {\mn@doi@[]}}
\def\mn@doi@[#1]#2{\def\@tempa{#1}\ifx\@tempa\@empty \href
  {http://dx.doi.org/#2} {doi:#2}\else \href {http://dx.doi.org/#2} {#1}\fi
  \endgroup}
\def\mn@eprint#1#2{\mn@eprint@#1:#2::\@nil}
\def\mn@eprint@arXiv#1{\href {http://arxiv.org/abs/#1} {{\tt arXiv:#1}}}
\def\mn@eprint@dblp#1{\href {http://dblp.uni-trier.de/rec/bibtex/#1.xml}
  {dblp:#1}}
\def\mn@eprint@#1:#2:#3:#4\@nil{\def\@tempa {#1}\def\@tempb {#2}\def\@tempc
  {#3}\ifx \@tempc \@empty \let \@tempc \@tempb \let \@tempb \@tempa \fi \ifx
  \@tempb \@empty \def\@tempb {arXiv}\fi \@ifundefined
  {mn@eprint@\@tempb}{\@tempb:\@tempc}{\expandafter \expandafter \csname
  mn@eprint@\@tempb\endcsname \expandafter{\@tempc}}}

\bibitem[\protect\citeauthoryear{{Asplund}, {Grevesse}, {Sauval}  \&
  {Scott}}{{Asplund} et~al.}{2009}]{Asplund2009}
{Asplund} M.,  {Grevesse} N.,  {Sauval} A.~J.,   {Scott} P.,  2009, \mn@doi
  [\araa] {10.1146/annurev.astro.46.060407.145222}, \href
  {https://ui.adsabs.harvard.edu/abs/2009ARA&A..47..481A} {47, 481}

\bibitem[\protect\citeauthoryear{{Barklem} et~al.,}{{Barklem}
  et~al.}{2005}]{Barklem2005}
{Barklem} P.~S.,  et~al., 2005, \mn@doi [\aap] {10.1051/0004-6361:20052967},
  \href {https://ui.adsabs.harvard.edu/abs/2005A&A...439..129B} {439, 129}

\bibitem[\protect\citeauthoryear{{Belczynski} et~al.,}{{Belczynski}
  et~al.}{2018}]{Belczynski2018}
{Belczynski} K.,  et~al., 2018, \mn@doi [\aap] {10.1051/0004-6361/201732428},
  \href {https://ui.adsabs.harvard.edu/abs/2018A&A...615A..91B} {615, A91}

\bibitem[\protect\citeauthoryear{{Bessell}}{{Bessell}}{2007}]{Bessell2007}
{Bessell} M.~S.,  2007, \mn@doi [\pasp] {10.1086/519981}, \href
  {https://ui.adsabs.harvard.edu/abs/2007PASP..119..605B} {119, 605}

\bibitem[\protect\citeauthoryear{{Castelli} \& {Kurucz}}{{Castelli} \&
  {Kurucz}}{2003}]{Castelli2003}
{Castelli} F.,  {Kurucz} R.~L.,  2003, in {Piskunov} N.,  {Weiss} W.~W.,
  {Gray} D.~F.,  eds,  Journal of Physics Conference Series Vol. 210, Modelling
  of Stellar Atmospheres. p.~A20 (\mn@eprint {arXiv} {astro-ph/0405087}),
  \mn@doi{10.48550/arXiv.astro-ph/0405087}

\bibitem[\protect\citeauthoryear{{Cayrel}}{{Cayrel}}{1988}]{Cayrel1988}
{Cayrel} R.,  1988, in {Cayrel de Strobel} G.,  {Spite} M.,  eds, ~ Vol. 132,
  The Impact of Very High S/N Spectroscopy on Stellar Physics. p.~345

\bibitem[\protect\citeauthoryear{{Cayrel} et~al.,}{{Cayrel}
  et~al.}{2004}]{Cayrel2004}
{Cayrel} R.,  et~al., 2004, \mn@doi [\aap] {10.1051/0004-6361:20034074}, \href
  {https://ui.adsabs.harvard.edu/abs/2004A&A...416.1117C} {416, 1117}

\bibitem[\protect\citeauthoryear{{Cescutti} \& {Chiappini}}{{Cescutti} \&
  {Chiappini}}{2014}]{Cescutti2014}
{Cescutti} G.,  {Chiappini} C.,  2014, \mn@doi [\aap]
  {10.1051/0004-6361/201423432}, \href
  {https://ui.adsabs.harvard.edu/abs/2014A&A...565A..51C} {565, A51}

\bibitem[\protect\citeauthoryear{{Chiti}, {Frebel}, {Ji}, {Jerjen}, {Kim}  \&
  {Norris}}{{Chiti} et~al.}{2018}]{Chiti2018}
{Chiti} A.,  {Frebel} A.,  {Ji} A.~P.,  {Jerjen} H.,  {Kim} D.,   {Norris}
  J.~E.,  2018, \mn@doi [\apj] {10.3847/1538-4357/aab4fc}, \href
  {https://ui.adsabs.harvard.edu/abs/2018ApJ...857...74C} {857, 74}

\bibitem[\protect\citeauthoryear{{Choplin} \& {Hirschi}}{{Choplin} \&
  {Hirschi}}{2020}]{Choplin2020}
{Choplin} A.,  {Hirschi} R.,  2020, in Journal of Physics Conference Series. p.
  012006 (\mn@eprint {arXiv} {2001.02341}),
  \mn@doi{10.1088/1742-6596/1668/1/012006}

\bibitem[\protect\citeauthoryear{{Choplin}, {Hirschi}, {Meynet}, {Ekstr{\"o}m},
  {Chiappini}  \& {Laird}}{{Choplin} et~al.}{2018}]{Choplin2018}
{Choplin} A.,  {Hirschi} R.,  {Meynet} G.,  {Ekstr{\"o}m} S.,  {Chiappini} C.,
   {Laird} A.,  2018, \mn@doi [\aap] {10.1051/0004-6361/201833283}, \href
  {https://ui.adsabs.harvard.edu/abs/2018A&A...618A.133C} {618, A133}

\bibitem[\protect\citeauthoryear{{Da Costa} et~al.,}{{Da Costa}
  et~al.}{2019}]{DaCosta2019}
{Da Costa} G.~S.,  et~al., 2019, \mn@doi [\mnras] {10.1093/mnras/stz2550},
  \href {https://ui.adsabs.harvard.edu/abs/2019MNRAS.489.5900D} {489, 5900}

\bibitem[\protect\citeauthoryear{{Da Costa}, {Bessell}, {Nordlander}, {Hughes},
  {Buder}, {Mackey}, {Spitler}  \& {Zucker}}{{Da Costa}
  et~al.}{2023}]{DaCosta2023}
{Da Costa} G.~S.,  {Bessell} M.~S.,  {Nordlander} T.,  {Hughes} A. C.~N.,
  {Buder} S.,  {Mackey} A.~D.,  {Spitler} L.~R.,   {Zucker} D.~B.,  2023,
  \mn@doi [\mnras] {10.1093/mnras/stad170}, \href
  {https://ui.adsabs.harvard.edu/abs/2023MNRAS.520..917D} {520, 917}

\bibitem[\protect\citeauthoryear{{Dekker}, {D'Odorico}, {Kaufer}, {Delabre}  \&
  {Kotzlowski}}{{Dekker} et~al.}{2000}]{Dekker2000}
{Dekker} H.,  {D'Odorico} S.,  {Kaufer} A.,  {Delabre} B.,   {Kotzlowski} H.,
  2000, in {Iye} M.,  {Moorwood} A.~F.,  eds,  Society of Photo-Optical
  Instrumentation Engineers (SPIE) Conference Series Vol. 4008, Optical and IR
  Telescope Instrumentation and Detectors. pp 534--545,
  \mn@doi{10.1117/12.395512}

\bibitem[\protect\citeauthoryear{{Ernandes}, {Castro}, {Barbuy}, {Spite},
  {Hill}, {Castilho}  \& {Evans}}{{Ernandes} et~al.}{2023}]{Ernandes2023}
{Ernandes} H.,  {Castro} M.~J.,  {Barbuy} B.,  {Spite} M.,  {Hill} V.,
  {Castilho} B.,   {Evans} C.~J.,  2023, \mn@doi [\mnras]
  {10.1093/mnras/stad1764}, \href
  {https://ui.adsabs.harvard.edu/abs/2023MNRAS.tmp.1712E} {}

\bibitem[\protect\citeauthoryear{Frebel \& Norris}{Frebel \&
  Norris}{2015}]{Frebel2015}
Frebel A.,  Norris J.~E.,  2015, \mn@doi [\araa]
  {10.1146/annurev-astro-082214-122423}, 53, 631

\bibitem[\protect\citeauthoryear{{Frischknecht} et~al.,}{{Frischknecht}
  et~al.}{2016}]{Frischknecht2016}
{Frischknecht} U.,  et~al., 2016, \mn@doi [\mnras] {10.1093/mnras/stv2723},
  \href {https://ui.adsabs.harvard.edu/abs/2016MNRAS.456.1803F} {456, 1803}

\bibitem[\protect\citeauthoryear{{Hirschi}}{{Hirschi}}{2007}]{Hirschi2007}
{Hirschi} R.,  2007, \mn@doi [\aap] {10.1051/0004-6361:20065356}, \href
  {https://ui.adsabs.harvard.edu/abs/2007A&A...461..571H} {461, 571}

\bibitem[\protect\citeauthoryear{{Ishigaki}, {Aoki}, {Arimoto}  \&
  {Okamoto}}{{Ishigaki} et~al.}{2014}]{Ishigaki2014}
{Ishigaki} M.~N.,  {Aoki} W.,  {Arimoto} N.,   {Okamoto} S.,  2014, \mn@doi
  [\aap] {10.1051/0004-6361/201322796}, \href
  {https://ui.adsabs.harvard.edu/abs/2014A&A...562A.146I} {562, A146}

\bibitem[\protect\citeauthoryear{{Jacobson} et~al.,}{{Jacobson}
  et~al.}{2015}]{Jacobson2015}
{Jacobson} H.~R.,  et~al., 2015, \mn@doi [\apj] {10.1088/0004-637X/807/2/171},
  \href {https://ui.adsabs.harvard.edu/abs/2015ApJ...807..171J} {807, 171}

\bibitem[\protect\citeauthoryear{{Ji}, {Frebel}, {Simon}  \& {Chiti}}{{Ji}
  et~al.}{2016}]{Ji2016}
{Ji} A.~P.,  {Frebel} A.,  {Simon} J.~D.,   {Chiti} A.,  2016, \mn@doi [\apj]
  {10.3847/0004-637X/830/2/93}, \href
  {https://ui.adsabs.harvard.edu/abs/2016ApJ...830...93J} {830, 93}

\bibitem[\protect\citeauthoryear{{Ji}, {Simon}, {Frebel}, {Venn}  \&
  {Hansen}}{{Ji} et~al.}{2019}]{Ji2019}
{Ji} A.~P.,  {Simon} J.~D.,  {Frebel} A.,  {Venn} K.~A.,   {Hansen} T.~T.,
  2019, \mn@doi [\apj] {10.3847/1538-4357/aaf3bb}, \href
  {https://ui.adsabs.harvard.edu/abs/2019ApJ...870...83J} {870, 83}

\bibitem[\protect\citeauthoryear{{Ji} et~al.,}{{Ji} et~al.}{2023}]{Ji2023}
{Ji} A.~P.,  et~al., 2023, \mn@doi [\aj] {10.3847/1538-3881/acad84}, \href
  {https://ui.adsabs.harvard.edu/abs/2023AJ....165..100J} {165, 100}

\bibitem[\protect\citeauthoryear{{Johnson}, {Herwig}, {Beers}  \&
  {Christlieb}}{{Johnson} et~al.}{2007}]{Johnson2007}
{Johnson} J.~A.,  {Herwig} F.,  {Beers} T.~C.,   {Christlieb} N.,  2007,
  \mn@doi [\apj] {10.1086/510114}, \href
  {https://ui.adsabs.harvard.edu/abs/2007ApJ...658.1203J} {658, 1203}

\bibitem[\protect\citeauthoryear{{J{\"o}nsson} et~al.,}{{J{\"o}nsson}
  et~al.}{2020}]{Jonsson2020}
{J{\"o}nsson} H.,  et~al., 2020, \mn@doi [\aj] {10.3847/1538-3881/aba592},
  \href {https://ui.adsabs.harvard.edu/abs/2020AJ....160..120J} {160, 120}

\bibitem[\protect\citeauthoryear{{Karakas} \& {Lattanzio}}{{Karakas} \&
  {Lattanzio}}{2007}]{Karakas2007}
{Karakas} A.,  {Lattanzio} J.~C.,  2007, \mn@doi [\pasa] {10.1071/AS07021},
  \href {https://ui.adsabs.harvard.edu/abs/2007PASA...24..103K} {24, 103}

\bibitem[\protect\citeauthoryear{{Keller} et~al.,}{{Keller}
  et~al.}{2014}]{Keller2014}
{Keller} S.~C.,  et~al., 2014, \mn@doi [\nat] {10.1038/nature12990}, \href
  {https://ui.adsabs.harvard.edu/abs/2014Natur.506..463K} {506, 463}

\bibitem[\protect\citeauthoryear{{Kobayashi}, {Ishigaki}, {Tominaga}  \&
  {Nomoto}}{{Kobayashi} et~al.}{2014}]{Kobayashi2014}
{Kobayashi} C.,  {Ishigaki} M.~N.,  {Tominaga} N.,   {Nomoto} K.,  2014,
  \mn@doi [\apjl] {10.1088/2041-8205/785/1/L5}, \href
  {https://ui.adsabs.harvard.edu/abs/2014ApJ...785L...5K} {785, L5}

\bibitem[\protect\citeauthoryear{{Kobayashi}, {Karakas}  \&
  {Lugaro}}{{Kobayashi} et~al.}{2020}]{Kobayashi2020}
{Kobayashi} C.,  {Karakas} A.~I.,   {Lugaro} M.,  2020, \mn@doi [\apj]
  {10.3847/1538-4357/abae65}, \href
  {https://ui.adsabs.harvard.edu/abs/2020ApJ...900..179K} {900, 179}

\bibitem[\protect\citeauthoryear{{Kobayashi} et~al.,}{{Kobayashi}
  et~al.}{2023}]{Kobayashi2023}
{Kobayashi} C.,  et~al., 2023, \mn@doi [\apjl] {10.3847/2041-8213/acad82},
  \href {https://ui.adsabs.harvard.edu/abs/2023ApJ...943L..12K} {943, L12}

\bibitem[\protect\citeauthoryear{{Marino} et~al.,}{{Marino}
  et~al.}{2019}]{Marino2019}
{Marino} A.~F.,  et~al., 2019, \mn@doi [\mnras] {10.1093/mnras/stz645}, \href
  {https://ui.adsabs.harvard.edu/abs/2019MNRAS.485.5153M} {485, 5153}

\bibitem[\protect\citeauthoryear{{Mashonkina} \& {Christlieb}}{{Mashonkina} \&
  {Christlieb}}{2014}]{Mashonkina2014}
{Mashonkina} L.,  {Christlieb} N.,  2014, \mn@doi [\aap]
  {10.1051/0004-6361/201423651}, \href
  {https://ui.adsabs.harvard.edu/abs/2014A&A...565A.123M} {565, A123}

\bibitem[\protect\citeauthoryear{{Mashonkina}, {Jablonka}, {Sitnova},
  {Pakhomov}  \& {North}}{{Mashonkina} et~al.}{2017}]{Mashonkina2017}
{Mashonkina} L.,  {Jablonka} P.,  {Sitnova} T.,  {Pakhomov} Y.,   {North} P.,
  2017, \mn@doi [\aap] {10.1051/0004-6361/201731582}, \href
  {https://ui.adsabs.harvard.edu/abs/2017A&A...608A..89M} {608, A89}

\bibitem[\protect\citeauthoryear{{McKenzie} et~al.,}{{McKenzie}
  et~al.}{2022}]{McKenzie2022}
{McKenzie} M.,  et~al., 2022, \mn@doi [\mnras] {10.1093/mnras/stac2254}, \href
  {https://ui.adsabs.harvard.edu/abs/2022MNRAS.516.3515M} {516, 3515}

\bibitem[\protect\citeauthoryear{{Meynet} \& {Maeder}}{{Meynet} \&
  {Maeder}}{2002}]{Meynet2002}
{Meynet} G.,  {Maeder} A.,  2002, \mn@doi [\aap] {10.1051/0004-6361:20020755},
  \href {https://ui.adsabs.harvard.edu/abs/2002A&A...390..561M} {390, 561}

\bibitem[\protect\citeauthoryear{{Meynet}, {Hirschi}, {Ekstrom}, {Maeder},
  {Georgy}, {Eggenberger}  \& {Chiappini}}{{Meynet} et~al.}{2010}]{Meynet2010}
{Meynet} G.,  {Hirschi} R.,  {Ekstrom} S.,  {Maeder} A.,  {Georgy} C.,
  {Eggenberger} P.,   {Chiappini} C.,  2010, \mn@doi [\aap]
  {10.1051/0004-6361/200913377}, \href
  {https://ui.adsabs.harvard.edu/abs/2010A&A...521A..30M} {521, A30}

\bibitem[\protect\citeauthoryear{Muggeo}{Muggeo}{2003}]{Muggeo2003}
Muggeo V.,  2003, \mn@doi [Statistics in medicine] {10.1002/sim.1545}, 22, 3055

\bibitem[\protect\citeauthoryear{{Neijssel} et~al.,}{{Neijssel}
  et~al.}{2019}]{Neijssel2019}
{Neijssel} C.~J.,  et~al., 2019, \mn@doi [\mnras] {10.1093/mnras/stz2840},
  \href {https://ui.adsabs.harvard.edu/abs/2019MNRAS.490.3740N} {490, 3740}

\bibitem[\protect\citeauthoryear{{Nidever} et~al.,}{{Nidever}
  et~al.}{2017}]{Nidever2017}
{Nidever} D.~L.,  et~al., 2017, \mn@doi [\aj] {10.3847/1538-3881/aa8d1c}, \href
  {https://ui.adsabs.harvard.edu/abs/2017AJ....154..199N} {154, 199}

\bibitem[\protect\citeauthoryear{{Nidever} et~al.,}{{Nidever}
  et~al.}{2020}]{Nidever2020}
{Nidever} D.~L.,  et~al., 2020, \mn@doi [\apj] {10.3847/1538-4357/ab7305},
  \href {https://ui.adsabs.harvard.edu/abs/2020ApJ...895...88N} {895, 88}

\bibitem[\protect\citeauthoryear{{Nomoto}, {Kobayashi}  \& {Tominaga}}{{Nomoto}
  et~al.}{2013}]{Nomoto2013}
{Nomoto} K.,  {Kobayashi} C.,   {Tominaga} N.,  2013, \mn@doi [\araa]
  {10.1146/annurev-astro-082812-140956}, \href
  {https://ui.adsabs.harvard.edu/abs/2013ARA&A..51..457N} {51, 457}

\bibitem[\protect\citeauthoryear{{Nordlander} et~al.,}{{Nordlander}
  et~al.}{2019}]{Nordlander2019}
{Nordlander} T.,  et~al., 2019, \mn@doi [\mnras] {10.1093/mnrasl/slz109}, \href
  {https://ui.adsabs.harvard.edu/abs/2019MNRAS.488L.109N} {488, L109}

\bibitem[\protect\citeauthoryear{{Norris} et~al.,}{{Norris}
  et~al.}{2013}]{Norris2013}
{Norris} J.~E.,  et~al., 2013, \mn@doi [\apj] {10.1088/0004-637X/762/1/28},
  \href {https://ui.adsabs.harvard.edu/abs/2013ApJ...762...28N} {762, 28}

\bibitem[\protect\citeauthoryear{{Oh}, {Nordlander}, {Da Costa}, {Bessell}  \&
  {Mackey}}{{Oh} et~al.}{2023}]{Oh2023}
{Oh} W.~S.,  {Nordlander} T.,  {Da Costa} G.~S.,  {Bessell} M.~S.,   {Mackey}
  A.~D.,  2023, \mn@doi [\mnras] {10.1093/mnras/stad1960}, \href
  {https://ui.adsabs.harvard.edu/abs/2023MNRAS.524..577O} {524, 577}

\bibitem[\protect\citeauthoryear{{Pignatari}, {Gallino}, {Heil}, {Wiescher},
  {K{\"a}ppeler}, {Herwig}  \& {Bisterzo}}{{Pignatari}
  et~al.}{2010}]{Pignatari2010}
{Pignatari} M.,  {Gallino} R.,  {Heil} M.,  {Wiescher} M.,  {K{\"a}ppeler} F.,
  {Herwig} F.,   {Bisterzo} S.,  2010, \mn@doi [\apj]
  {10.1088/0004-637X/710/2/1557}, \href
  {https://ui.adsabs.harvard.edu/abs/2010ApJ...710.1557P} {710, 1557}

\bibitem[\protect\citeauthoryear{Pilgrim}{Pilgrim}{2021}]{Pilgrim2021}
Pilgrim C.,  2021, \mn@doi [Journal of Open Source Software]
  {10.21105/joss.03859}, 6, 3859

\bibitem[\protect\citeauthoryear{{Placco}, {Frebel}, {Beers}  \&
  {Stancliffe}}{{Placco} et~al.}{2014}]{Placco2014}
{Placco} V.~M.,  {Frebel} A.,  {Beers} T.~C.,   {Stancliffe} R.~J.,  2014,
  \mn@doi [\apj] {10.1088/0004-637X/797/1/21}, \href
  {https://ui.adsabs.harvard.edu/abs/2014ApJ...797...21P} {797, 21}

\bibitem[\protect\citeauthoryear{{Reggiani}, {Schlaufman}, {Casey}, {Simon}  \&
  {Ji}}{{Reggiani} et~al.}{2021}]{Reggiani2021}
{Reggiani} H.,  {Schlaufman} K.~C.,  {Casey} A.~R.,  {Simon} J.~D.,   {Ji}
  A.~P.,  2021, \mn@doi [\aj] {10.3847/1538-3881/ac1f9a}, \href
  {https://ui.adsabs.harvard.edu/abs/2021AJ....162..229R} {162, 229}

\bibitem[\protect\citeauthoryear{{Simmerer}, {Sneden}, {Cowan}, {Collier},
  {Woolf}  \& {Lawler}}{{Simmerer} et~al.}{2004}]{Simmerer2004}
{Simmerer} J.,  {Sneden} C.,  {Cowan} J.~J.,  {Collier} J.,  {Woolf} V.~M.,
  {Lawler} J.~E.,  2004, \mn@doi [\apj] {10.1086/424504}, \href
  {https://ui.adsabs.harvard.edu/abs/2004ApJ...617.1091S} {617, 1091}

\bibitem[\protect\citeauthoryear{{Sk{\'u}lad{\'o}ttir}, {Hansen}, {Salvadori}
  \& {Choplin}}{{Sk{\'u}lad{\'o}ttir} et~al.}{2019}]{Skuladottir2019}
{Sk{\'u}lad{\'o}ttir} {\'A}.,  {Hansen} C.~J.,  {Salvadori} S.,   {Choplin} A.,
   2019, \mn@doi [\aap] {10.1051/0004-6361/201936125}, \href
  {https://ui.adsabs.harvard.edu/abs/2019A&A...631A.171S} {631, A171}

\bibitem[\protect\citeauthoryear{{Sk{\'u}lad{\'o}ttir}
  et~al.,}{{Sk{\'u}lad{\'o}ttir} et~al.}{2021}]{Skuladottir2021}
{Sk{\'u}lad{\'o}ttir} {\'A}.,  et~al., 2021, \mn@doi [\apjl]
  {10.3847/2041-8213/ac0dc2}, \href
  {https://ui.adsabs.harvard.edu/abs/2021ApJ...915L..30S} {915, L30}

\bibitem[\protect\citeauthoryear{{Sneden} \& {Cowan}}{{Sneden} \&
  {Cowan}}{2003}]{Sneden2003}
{Sneden} C.,  {Cowan} J.~J.,  2003, \mn@doi [Science]
  {10.1126/science.1077506}, \href
  {https://ui.adsabs.harvard.edu/abs/2003Sci...299...70S} {299, 70}

\bibitem[\protect\citeauthoryear{{Spite}, {Spite}, {Fran{\c{c}}ois},
  {Bonifacio}, {Caffau}  \& {Salvadori}}{{Spite} et~al.}{2018}]{Spite2018}
{Spite} M.,  {Spite} F.,  {Fran{\c{c}}ois} P.,  {Bonifacio} P.,  {Caffau} E.,
  {Salvadori} S.,  2018, \mn@doi [\aap] {10.1051/0004-6361/201833548}, \href
  {https://ui.adsabs.harvard.edu/abs/2018A&A...617A..56S} {617, A56}

\bibitem[\protect\citeauthoryear{{Tafelmeyer} et~al.,}{{Tafelmeyer}
  et~al.}{2010}]{Tafelmeyer2010}
{Tafelmeyer} M.,  et~al., 2010, \mn@doi [\aap] {10.1051/0004-6361/201014733},
  \href {https://ui.adsabs.harvard.edu/abs/2010A&A...524A..58T} {524, A58}

\bibitem[\protect\citeauthoryear{{Yong} et~al.,}{{Yong}
  et~al.}{2013}]{Yong2013}
{Yong} D.,  et~al., 2013, \mn@doi [\apj] {10.1088/0004-637X/762/1/26}, \href
  {https://ui.adsabs.harvard.edu/abs/2013ApJ...762...26Y} {762, 26}

\bibitem[\protect\citeauthoryear{{Yong} et~al.,}{{Yong}
  et~al.}{2021a}]{Yong2021}
{Yong} D.,  et~al., 2021a, \mn@doi [\mnras] {10.1093/mnras/stab2001}, \href
  {https://ui.adsabs.harvard.edu/abs/2021MNRAS.507.4102Y} {507, 4102}

\bibitem[\protect\citeauthoryear{{Yong} et~al.,}{{Yong}
  et~al.}{2021b}]{Yong2021b}
{Yong} D.,  et~al., 2021b, \mn@doi [\nat] {10.1038/s41586-021-03611-2}, \href
  {https://ui.adsabs.harvard.edu/abs/2021Natur.595..223Y} {595, 223}

\makeatother
\end{thebibliography}

%\begin{thebibliography}{99}
%\end{thebibliography}

%%%%%%%%%%%%%%%%%%%%%%%%%%%%%%%%%%%%%%%%%%%%%%%%%%

%%%%%%%%%%%%%%%%% APPENDICES %%%%%%%%%%%%%%%%%%%%%

\begin{landscape}
\begin{table}
	\centering
	\caption{Abundance table showing the abundance measurement for each star. The measurement uncertainties} are given for each element.
	\label{tab:test}
	\begin{tabular}{lcccccccccccccccccccccccccccc} % four columns, alignment for each
		\hline 
		SMSS DR3 &
		[Fe/H] &
		$\rm{[C/Fe]_{raw}}$ &
            $\rm{[C/Fe]_{corr}}$ &
            [N/Fe] & 
	    [O/Fe] & 
            [Na/Fe] &  
            [Mg/Fe] &  
            [Al/Fe] &  
            [Si/Fe] &
            [Ca/Fe] \\ &
            [Sc/Fe]  & 
            [\ion{Ti}{I}/Fe] & 
            [\ion{Ti}{II}/Fe] &
            [Cr/Fe] &
            [Mn/Fe] &
            [\ion{Fe}{II}/H] &
            [Co/Fe] &
            [Ni/Fe] &
            [Zn/Fe] &
            [Sr/Fe] \\ &
            [Y/Fe] &
            [Zr/Fe] & 
            [Ba/Fe] & 
            [La/Fe] &
            [Nd/Fe] &
            [Sm/Fe] &
            [Eu/Fe] &
            [Dy/Fe]\\
		\hline
%$\pm 0.05$. 
500287810 & 
$-2.49 \pm0.11$ & 
$-0.80 \pm0.15$ & 
$-0.12 \pm0.15$ & 
$<0.50$ & 
$0.85 \pm0.12$ & 
$-0.06 \pm0.13$ & 
$0.22 \pm0.06$ & 
$-0.24 \pm0.19$ & 
$0.29 \pm0.12$ &
$0.16 \pm0.05$ \\ & 
$0.07 \pm0.19$  &
$-0.02 \pm0.08$ & 
$0.12 \pm0.19$ & 
$-0.31 \pm0.12$ & 
$-0.42 \pm0.26$ & 
$-2.73 \pm0.11$ & 
$0.25 \pm0.17$ & 
$-0.05 \pm0.05$ & 
$0.00 \pm0.06$ & 
$-0.88 \pm0.19$ \\ &
$-0.68 \pm0.23$ &
$-0.39 \pm0.07$ & 
$-0.76 \pm0.19$ & 
$<-0.85$ & 
$<-0.65$ & 
$....$ & 
$<-0.45$ & 
$<-0.40$ \\ 
& \\

500382880 & 
$-2.65 \pm0.10$ & 
$-0.60 \pm0.19$ & 
$0.04 \pm0.19$ &
$<0.5$ & 
$1.10 \pm0.27$ & 
$0.15 \pm0.12$ & 
$0.24 \pm0.05$ & 
$-0.07 \pm0.16$ & 
$0.13 \pm0.13$ & 
$0.26 \pm0.05$ \\ &
$0.30 \pm0.17$ &
$0.25 \pm0.05$ &
$0.39 \pm0.16$ & 
$-0.24 \pm0.09$ & 
$-0.31 \pm0.21$ & 
$-2.61 \pm0.10$ & 
$0.16 \pm0.11$ & 
$-0.24 \pm0.17$ & 
$0.10 \pm0.05$ & 
$-0.08 \pm0.15$ \\ & 
$-0.10 \pm0.16$ &
$0.23 \pm0.05$ &
$0.02 \pm0.17$ & 
$0.28 \pm0.07$ & 
$0.43 \pm0.08$ & 
$0.60 \pm0.09$ & 
$0.80 \pm0.08$ & 
$0.85 \pm0.08$ & \\
& \\

497682788 & 
$-2.80 \pm0.10$ & 
$-0.90 \pm0.13$ & 
$-0.26 \pm0.13$ & 
$<0.70$ & 
$1.05 \pm0.16$ & 
$-0.03 \pm0.09$ & 
$0.33 \pm0.08$ & 
$-0.54 \pm0.15$ & 
$0.53 \pm0.10$ &
$0.27 \pm0.08$ \\ &
$0.22 \pm0.18$ & 
$0.15 \pm0.04$ & 
$0.46 \pm0.22$ & 
$-0.28 \pm0.05$ & 
$-0.49 \pm0.17$ & 
$-2.65 \pm0.10$ & 
$0.14 \pm0.08$ & 
$-0.11 \pm0.15$ & 
$0.35 \pm0.04$ & 
$-0.50 \pm0.18$ \\ &
$-0.23 \pm0.20$ & 
$0.19 \pm0.05$ & 
$0.08 \pm0.19$ & 
$0.50 \pm0.06$ & 
$0.73 \pm0.06$ & 
$0.80 \pm0.08$ & 
$1.03 \pm0.06$ & 
$1.00 \pm0.08$ \\ 
& \\

471915910 & 
$-2.90 \pm0.07$ & 
$-0.35 \pm0.26$ & 
$0.28 \pm0.26$ & 
$<0.90$ & 
$....$ & 
$0.15 \pm0.14$ & 
$0.41 \pm0.06$ & 
$-0.09 \pm0.20$ & 
$0.66 \pm0.15$ &
$0.41 \pm0.06$ \\ &
$0.24 \pm0.16$ & 
$0.30 \pm0.06$ & 
$0.38 \pm0.16$ & 
$-0.18 \pm0.08$ & 
$-0.48 \pm0.18$ & 
$-2.83 \pm0.07$ & 
$0.23 \pm0.09$ & 
$-0.19 \pm0.16$ & 
$0.35 \pm0.11$ & 
$-0.10 \pm0.18$ \\ &
$0.10 \pm0.18$ &
$0.26 \pm0.12$ & 
$-0.40 \pm0.14$ & 
$<-0.25$ & 
$<-0.05$ & 
$....$ & 
$<0.3$ & 
$<0.25$ \\ 		
& \\

500766372 & 
$-2.90 \pm0.08$ & 
$-0.55 \pm0.27$ & 
$0.04 \pm0.27$ & 
$<0.90$ & 
$....$ & 
$0.19 \pm0.09$ & 
$-0.02 \pm0.09$ & 
$-0.45 \pm0.14$ & 
$0.50 \pm0.12$ &
$0.12 \pm0.08$ \\ &
$-0.30 \pm0.13$ & 
$0.25 \pm0.06$ & 
$0.22 \pm0.15$ & 
$-0.17 \pm0.05$ & 
$-0.07 \pm0.15$ & 
$-2.82 \pm0.08$ & 
$0.28 \pm0.07$ & 
$0.09 \pm0.13$ & 
$0.75 \pm0.04$ & 
$-1.30 \pm0.12$ \\ &
$< -1.00$ & 
$-0.33 \pm0.08$ & 
$-1.21 \pm0.17$ & 
$<-0.20$ & 
$<0.15$ & 
$....$ & 
$<-0.10$ & 
$<-0.50$ \\ 
& \\

497519424 & 
$-3.13 \pm0.08$ & 
$0.00 \pm0.14$ & 
$0.63 \pm0.14$ & 
$1.70 \pm0.11$ & 
$....$ & 
$1.25 \pm0.15$ & 
$0.43 \pm0.05$ & 
$-0.10 \pm0.21$ & 
$0.17 \pm0.19$ &
$0.30 \pm0.05$ \\ &
$0.22 \pm0.17$ & 
$0.24 \pm0.04$ & 
$0.50 \pm0.15$ & 
$-0.33 \pm0.05$ & 
$-0.84 \pm0.12$ & 
$-3.06 \pm0.08$ & 
$0.32 \pm0.09$ & 
$-0.10 \pm0.24$ & 
$0.35 \pm0.05$ & 
$-0.59 \pm0.17$ \\ &
$-0.28 \pm0.20$ & 
$0.15 \pm0.05$ & 
$-0.70 \pm0.15$ & 
$<-0.20$ & 
$<0.00$ & 
$....$ & 
$<0.15$ & 
$<0.15$ \\ 
& \\

499901368 & 
$-3.13 \pm0.09$ & 
$-0.10 \pm0.23$ & 
$0.51 \pm0.23$ & 
$<1.10$ & 
$....$ & 
$0.21 \pm0.05$ & 
$0.34 \pm0.06$ & 
$-0.42 \pm0.12$ & 
$1.03 \pm0.07$ &
$0.25 \pm0.08$ \\ &
$0.24 \pm0.19$ &
$0.21 \pm0.11$ & 
$0.26 \pm0.17$ & 
$-0.18 \pm0.05$ & 
$-0.10 \pm0.29$ & 
$-3.14 \pm0.09$ & 
$0.35 \pm0.04$ & 
$-0.06 \pm0.19$ & 
$0.40 \pm0.09$ & 
$-0.55 \pm0.21$ \\ &
$-0.23 \pm0.16$ & 
$0.14 \pm0.07$ & 
$-0.85 \pm0.20$ & 
$<0.10$ & 
$<0.20$ & 
$....$ & 
$<0.30$ & 
$<0.10$ \\ 

		\hline
	\end{tabular}
	\end{table}
\end{landscape}

%%%%%%%%%%%%%%%%%%%%%%%%%%%%%%%%%%%%%%%%%%%%%%%%%%

% Don't change these lines
\bsp	% typesetting comment
\label{lastpage}
\end{document}